\documentclass[prx,twocolumn,amsmath,amssymb]{revtex4} 

\usepackage{color}
\usepackage{bm}
\usepackage{dcolumn}
\usepackage[pdftex]{graphicx}    

\usepackage{slashed}
\usepackage{amsmath}\usepackage{accents}

\newcommand{\1}{\mbox{1}\hspace{-0.25em}\mbox{l}}

\begin{document}

\title{
Moir\'e Landau levels of a $C_4$-symmetric twisted bilayer system 
in the absence of a magnetic field
}

\author{Yuki Soeda, Koichi Asaga, and  Takahiro Fukui}
\affiliation{Department of Physics, Ibaraki University, Mito 310-8512, Japan}

\date{\today}

\begin{abstract}
It is widely known that the twisted bilayer graphene (TBG) shows flat bands at magic angles, 
which can be well described  by the effective continuum model derived by Bistritzer and MacDonald (BM).   
We propose in this paper
a similar twisted bilayer system but defined on the square lattice with $\pi$ flux per plaquette, and 
study its spectrum  using the BM Hamiltonian with a mass term which is originated from the staggered potential.  
The basic difference between the TBG and the present model
is simply rotational symmetry, $C_3$ versus $C_4$, as well as a mass term.
Nevertheless, the feature of the flat bands is quite different: those of the TBG appear at magic angles only,
while the present model shows many flat bands, which are reminiscent of Landau levels,
quite stably at any angles even in the absence of
a magnetic field  other than $\pi$ flux, which keeps time reversal (TR) symmetry.
Moreover,  flat bands emerge in the mass gap of the Dirac spectrum, and each state 
composing these flat bands is well-localized at the position forming the moir\'e lattice. 
It turns out that the moir\'e potential serves as a periodic magnetic field, 
which can give  energies smaller that the gap around moir\'e lattice positions.
We derive a local Hamiltonian valid around the moir\'e lattice sites and show that it indeed reproduces the energies of 
the flat bands within the mass gap. 
Since these mid-gap states are localized at the moir\'e lattice, they form degenerate levels, which may be referred to 
as moir\'e Landau levels, although the mechanism of degeneracies are different from the conventional Landau levels.
Interestingly, doubled fermions of the BH Hamiltonian associated with two layers have opposite charges when they couple with the 
effective moir\'e magnetic filed, which
concern TR symmetry.
This is a generic feature of the $C_4$ symmetric moir\'e system described by the BM Hamiltonian.
\end{abstract}

\pacs{
}

\maketitle

\section{Introduction}

Flat bands provide a promising platform for studying strongly correlated systems.
Recent discovery of flat bands in the twisted bilayer graphene (TBG) \cite{Lopes-dos-Santos:2007aa,Suarez-Morell:2010tt,Mele:2011ts,Bistritzer:2011ab,Moon:2012ui,Lopes-dos-Santos:2012vk,Moon:2013vv}
has been attracting much current interest, in which 
superconductivity \cite{Fatemi:2018aa,Cao:2018aa,Lee:2019uv}, 
correlated insulating phase \cite{Cao:2018tp,Zhang:2019aa,Lee:2019uv}, 
and nematic behavior  \cite{Yuan:2021wg} 
have been observed experimentally.
Symmetries and associated topological properties of the TBG has been extensively studied \cite{Zou:2018aa,Ahn:2019ab,Hejazi:2019aa,Song:2019wo,Bultinck:2020wq}.
The origin of the flat bands in TBG has been clarified in \cite{Tarnopolsky:2019aa,Wang:2021up},
where chiral symmetry as well as threefold rotational  $C_3$ symmetry play a crucial role.
Since moir\'e structure induces a long-period moir\'e lattice with the same symmetry as the microscopic lattice, 
the possibility of revealing the electronic structure of solids under an extremely 
strong magnetic field has been suggested \cite{Bistritzer:2011aa,Koshino:2015um,Zhang:2019tt,Hejazi:2019vr,Crosse:2020tp,Sheffer:2021wi}. 
Indeed, the TBG under $2\pi$ flux per moir\'e plaquette has recently been observed 
\cite{das2021observation,herzogarbeitman2021reentrant}.
The flat bands of the TBG occurs even in generic multilayer systems: 
The relationship between magic angles and the number of layers has been conjectured in Ref. \cite{Khalaf:2019ud}.
Detailed discussions including above can be found in the series of papers 
\cite{Bernevig:2021tu,Song:2021ti,Bernevig:2021vj,Lian:2021tm,Bernevig:2021wa,Xie:2021ty}
and a review
\cite{ledwith2021strong}.

These analyses are based on the
Bistritzer and MacDonald (BM) Hamiltonian 
composed of the doubled massless Dirac fermions with SO(2) rotational symmetry
which is broken to $C_3$ due to moir\'e potentials. 
Despite its simple structure, it offers rich physics aforementioned once interactions are introduced.
In order to further study the universal behavior of the BM Hamiltonian and unexpected phenomena behind it, 
it may be interesting to address the question whether flat bands occur in $C_n$ symmetric potentials.
Such a generic BM Hamiltonian can be constructed solely by the symmetry argument based on doubled 
Dirac fermions, while it is also known that
the honeycomb lattice can be topologically deformed into the square lattice with $\pi$ flux per plaquette 
with keeping linear dispersions around the band center \cite{Hatsugai:2006aa}. 
This motivates us to investigate a moir\'e system with $C_4$ symmetry based on the $\pi$-flux model.




In this paper, we consider a free fermion model defined on the square lattice with $\pi$ flux per plaquette,  which allows,
similarly to the graphene, the Dirac fermions close to zero energy.
This model has not only fourfold rotational $C_4$ symmetry but also time reversal (TR) symmetry 
because of the special magnetic field.
A staggered potential opens a mass gap in the Dirac fermion spectrum and breaks TR symmetry.
We derive an effective continuum model composed of doubled massive Dirac fermions
for the twisted bilayer $\pi$-flux models, according to BM \cite{Bistritzer:2011ab}.
Such an effective theory is universal in that it does not depend on the details of the lattice model:
In fact, for the single-layer system, 
the graphene as well as the $\pi$-flux model are described by the same Dirac Hamiltonian around the band center.

However, for the twisted bilayer systems, interlayer couplings are reflected by the rotational symmetries of the 
lattice model. 
Namely, the Dirac model of the present system has $C_4$ symmetry, which is in sharp contrast to $C_3$ symmetry 
of the TBG.
It then turns out that this difference of rotational symmetry induces much more differences in the spectrum:
The TBG shows nearly-flat bands solely at magic angles, whereas {\it the present model shows 
flat bands stably at any small angles}.
Remarkably, the latter look like Landau levels even in the absence of a magnetic field other than $\pi$ flux.
Indeed, the model has time reversal (TR) symmetry which is broken solely by the staggered potential.
Such flat bands are revealed to be states localized at the moir\'e lattice, which can be described simply by
doubled Dirac fermions with opposite charges in a uniform magnetic field.

This paper is organized as follows.
In Sec. \ref{s:single}, we define the  model on the square lattice, where we take a specific gauge representing $\pi$ flux
per plaquette in order to make $C_4$ symmetry manifest.
Based on this lattice model, we derive the continuum Dirac Hamiltonian.
Section \ref{s:twist} is devoted to the construction of the twisted system. 
Namely, we derive the effective Hamiltonian of the BM type, by requiring $C_4$ symmetry as well as 
calculating the interlayer coupling.
Then, numerical calculations of the effective Dirac Hamiltonian thus obtained shows quite characteristic band structure:
Many flat bands appear within the mass gap which remind us of the Landau levels  even though
the present system includes no magnetic field other than $\pi$ flux per plaquette. 
The density profiles of those flat bands reveal that each state of the flat bands is localized at each moir\'e lattice. 
To clarify the nature of such a localized state, we derive, 
in Sec. \ref{s:landau},  an effective Hamiltonian valid around  zero energy within the mass gap.
Remarkably, such a Hamiltonian is nothing but the Dirac Hamiltonian in the presence of a uniform magnetic field.
Nevertheless, TR invariance is guaranteed in the massless case, since 
the Dirac Hamiltonian includes doubled fermions with opposite effective charges.
It turns out that the energies of the flat bands is indeed well reproduced  by such an effective Dirac Hamiltonian.
In sec. \ref{s:summary}, we give the summary and discussion, especially on the experimental feasibility.
One of candidates for the moir\'e landau levels proposed in this paper is  
a symmetry-protected Dirac semimetal with $C_4$ symmetry, 
in which a mass gap is needed by  adding some symmetry-breaking  perturbations.
Another candidate which has more intimate relationship with the $\pi$-flux model is also proposed.
Generically, a magnetic field which gives $\pi$ flux per plaquette is too large to be realized in experiments.
Therefore, by the use of an enlarged long-periodic modulation on the lattice, we propose a model in a uniform magnetic field
giving  $\pi$ flux per unit cell composed of many sites. We demonstrate that one example of such systems indeed 
shows a doubled Dirac dispersion in the Brillouin zone. 

\section{Basic single layer model}\label{s:single}

\begin{figure}[h]
\begin{center}
\begin{tabular}{c}
\includegraphics[width=0.7\linewidth]{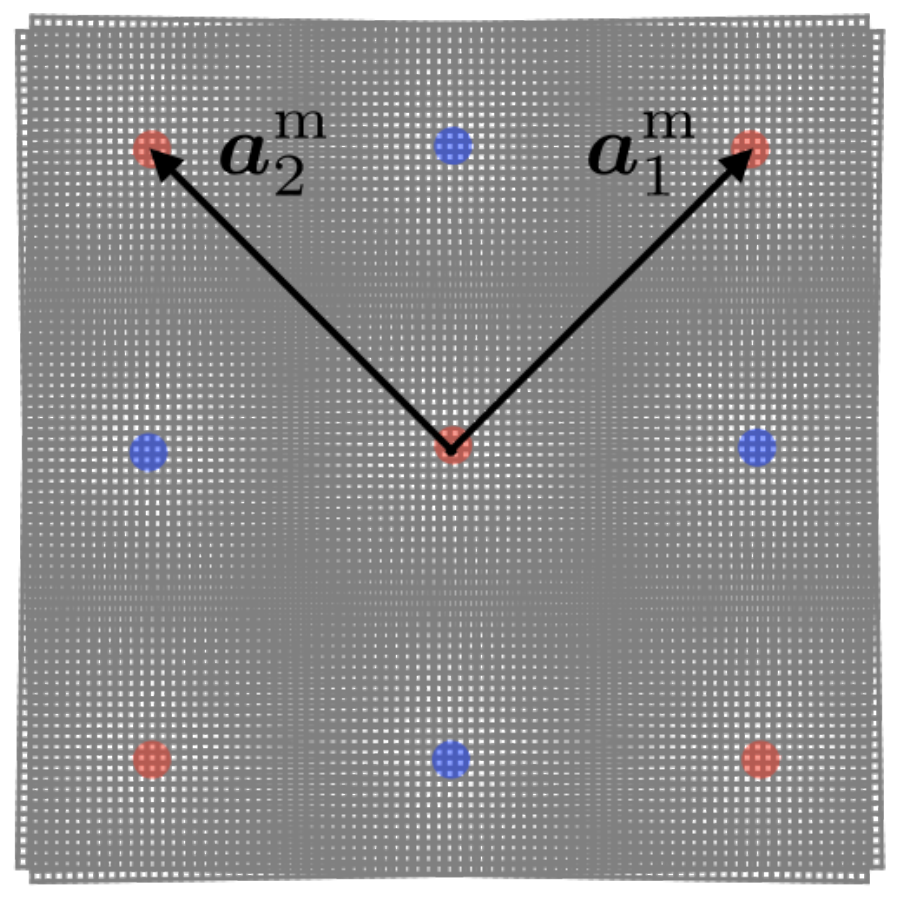}
\end{tabular}
\caption{
An example of the moir\'e pattern of the twisted bilayer system on the square lattice.
Small regions colored by red and blue stand for the $AA$ and $AB$ stack points, respectively.
The arrows marked by $\bm a_1^{\rm m}$ and  $\bm a_2^{\rm m}$ are primitive moir\'e translation vectors.
See the text.
}
\label{f:moire}
\end{center}
\end{figure}

There are several attempts to consider wider class of the moir\'e pattern in various 
Bravais lattices \cite{Akashi:2017tk,Kariyado:2019ub}.
This paper, however, aims at revealing  the universal behavior of the BM model, so that 
we start with a model showing linear dispersions.
To this end, we introduce, in this section, a well-known example, the $\pi$-flux model defined on the square lattice,
which can be regarded as  topological deformation of the graphene \cite{Hatsugai:2006aa}.
As can be seen in Fig. \ref{f:moire}, the square lattice shows the moir\'e pattern in which the $C_4$ symmetry 
is clearly seen. 
Thus, in this section,  we define the $\pi$-flux model  in a specific gauge with clear $C_4$ symmetry.

\subsection{Lattice model} \label{s:lattice_single}

\begin{figure}[h]
\begin{center}
\begin{tabular}{c}
\includegraphics[width=0.9\linewidth]{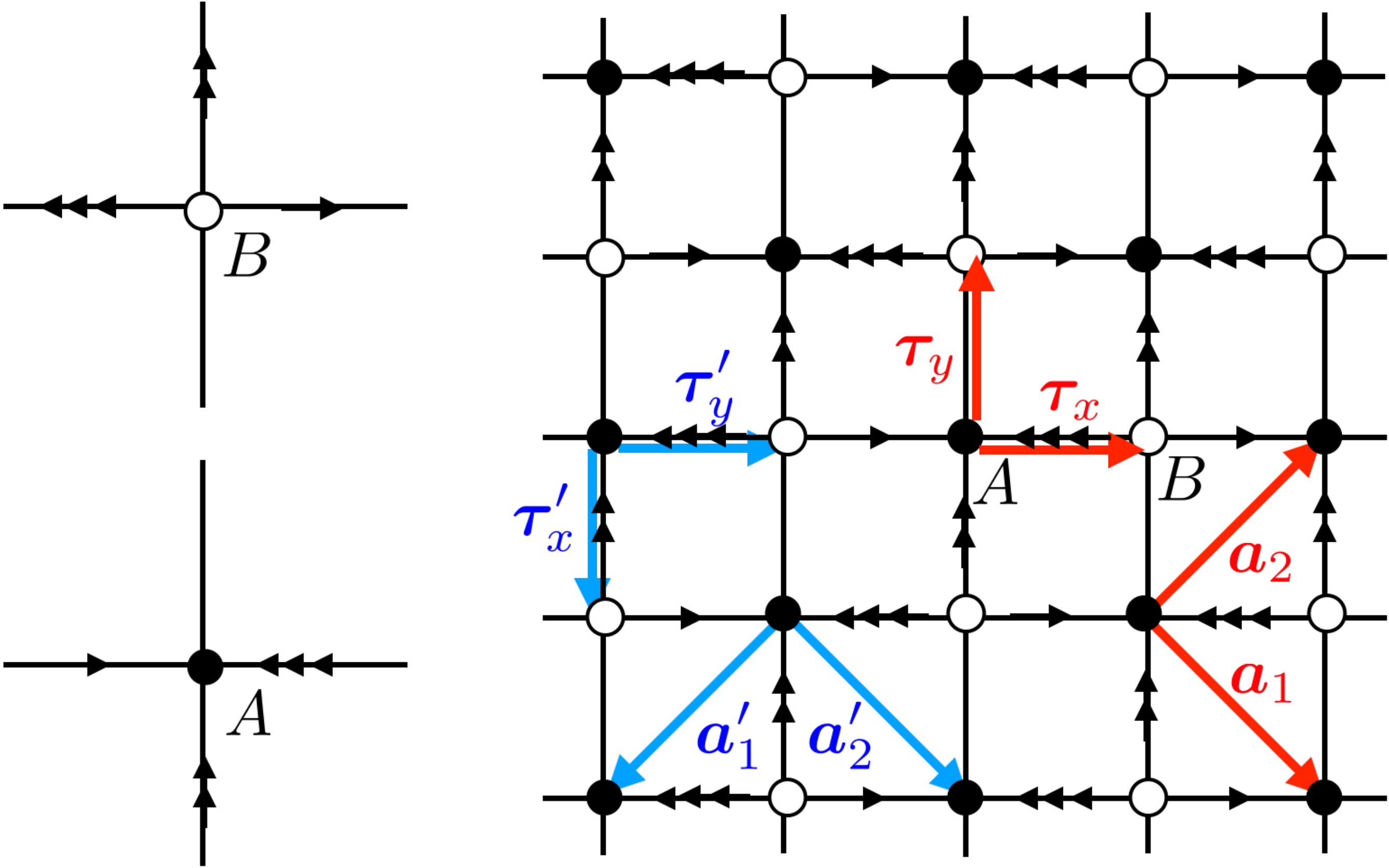}
\end{tabular}
\caption{
Each arrow on the bond denotes the phase of the hopping, $i$, and hence, the double and triple arrow means $-1$ and 
$-i$, respectively.
The blue arrows are for the $\pi/2$-rotated system.
}
\label{f:lat}
\end{center}
\end{figure}

We introduce a model with a nearest-neighbor hopping $t_{AB}$  as well as an onsite staggered potential $\pm\Delta$.
We assume that  the nearest-neighbor hopping 
from $B$ at $\bm r_B$ to $A$ at $\bm r_A$
is  complex in general, dependent on their relative position $\bm r_A-\bm r_B=r(\cos\theta,\sin\theta)$ such that
$t_{AB}(\bm r_A-\bm r_B)=te^{i(\theta+\theta_0)}$, where we choose $\theta_0=\pi/2$ for later convenience.
As can be seen in Fig. \ref{f:lat}, the total amount of phases around each plaquette is just $\pm\pi$,
implying that the present model is equivalent to the $\pi$-flux model.

The primitive translation  vectors are $\bm a_1=a(1,-1)$ and $\bm a_2=a(1,1)$, and nearest-neighbor sites are 
connected by $\bm \tau_x=a(1,0)=(\bm a_1+\bm a_2)/2$ and $\bm \tau_y=a(0,1)=(-\bm a_1+\bm a_2)/2$, where $a$ is the lattice constant.
The Hamiltonian is then written as
\begin{alignat}1
H
&=-t\sum_{\bm r}c_{A,\bm r}^\dagger\left[
-i(\delta_x-\delta^*_x)+(\delta_y-\delta^*_y)\right]c_{B,\bm r}
+{\rm H.c.}
\nonumber\\&
+\Delta\sum_{\bm r}[c_{A,\bm r}^\dagger c_{A,\bm r}-\delta_x(c_{B,\bm r}^\dagger c_{B,\bm r})],
\label{HamPi}
\end{alignat}
where the forward and backward shift operators are defined by 
$\delta_jf_{\bm r}=f_{\bm r+\bm\tau_j}$ and $\delta^*_jf_{\bm r}=f_{\bm r-\bm\tau_j}$.
Note here that $\delta_j-\delta^*_j=\nabla_j+\nabla_j^*$, where $\nabla_j$ and $\nabla_j^*$ are
the forward and backward difference operators, so that the Hamiltonian (\ref{HamPi}) is nothing but the
massive Dirac Hamiltonian on the lattice.

Let us consider the $\pi/2$ rotation of the system. 
This is equivalent to the $-\pi/2$ rotation of the axes, so that 
let us introduce the following vectors denoted by blue arrows in Fig. \ref{f:lat},
\begin{alignat}1
&\bm a'_1=-\bm a_2,\quad \bm a'_2=\bm a_1;\quad
\bm\tau_x'=-\bm\tau_y,\quad \bm\tau_y'=\bm\tau_x,
\label{TraLawTau}
\end{alignat} 
Also, we introduce the gauge transformation $c_{A,\bm r}=c_{A,\bm r}'$ and $c_{B,\bm r}=ic_{B,\bm r}'$, namely,
in the matrix notation, $\bm c_{\bm r}=W\bm c'_{\bm r}$ with 
\begin{alignat}1
W=\begin{pmatrix}
1&\\&i
\end{pmatrix}
=\omega\begin{pmatrix}
\bar\omega&\\&\omega
\end{pmatrix},\quad (\omega=e^{i\pi/4}),
\label{GauTra}
\end{alignat}
where we have defined $\bm c_{\bm r}\equiv (c_{A,\bm r},c_{B,\bm r})^T$.
Then, in the rotated basis,  the Hamiltonian is given by 
\begin{alignat}1
H&=-t\sum_{\bm r}{c}_{A,\bm r}^\dagger \left[
-i(\delta'_y-\delta'^*_y)-(\delta'_x-\delta'^*_x)
\right]c_{B,\bm r}+{\rm H.c.}
\nonumber\\
&+\Delta\sum_{\bm r}[c_{A,\bm r}^\dagger c_{A,\bm r}-\delta_y'(c_{B,\bm r}^\dagger c_{B,\bm r})]
\nonumber\\
&=-t\sum_{\bm r}{c}_{A,\bm r}'^\dagger \left[
-i(\delta'_x-\delta'^*_x)+(\delta'_y-\delta'^*_y)
\right]c_{B,\bm r}'+{\rm H.c.}
\nonumber\\
&+\Delta\sum_{\bm r}[c_{A,\bm r}'^\dagger c_{A,\bm r}'-\delta_x'(c_{B,\bm r}'^\dagger c_{B,\bm r}')],
\label{LatC4}
\end{alignat}
where we have used the fact that 
$\sum_{\bm r}\delta_x(\cdots)=\sum_{\bm r}\delta_y(\cdots)$.
Thus  Eq. (\ref{LatC4}) implies the invariance of the Hamiltonian under $\pi/2$ rotation.

\begin{figure}[h]
\begin{center}
\begin{tabular}{c}
\includegraphics[width=0.97\linewidth]{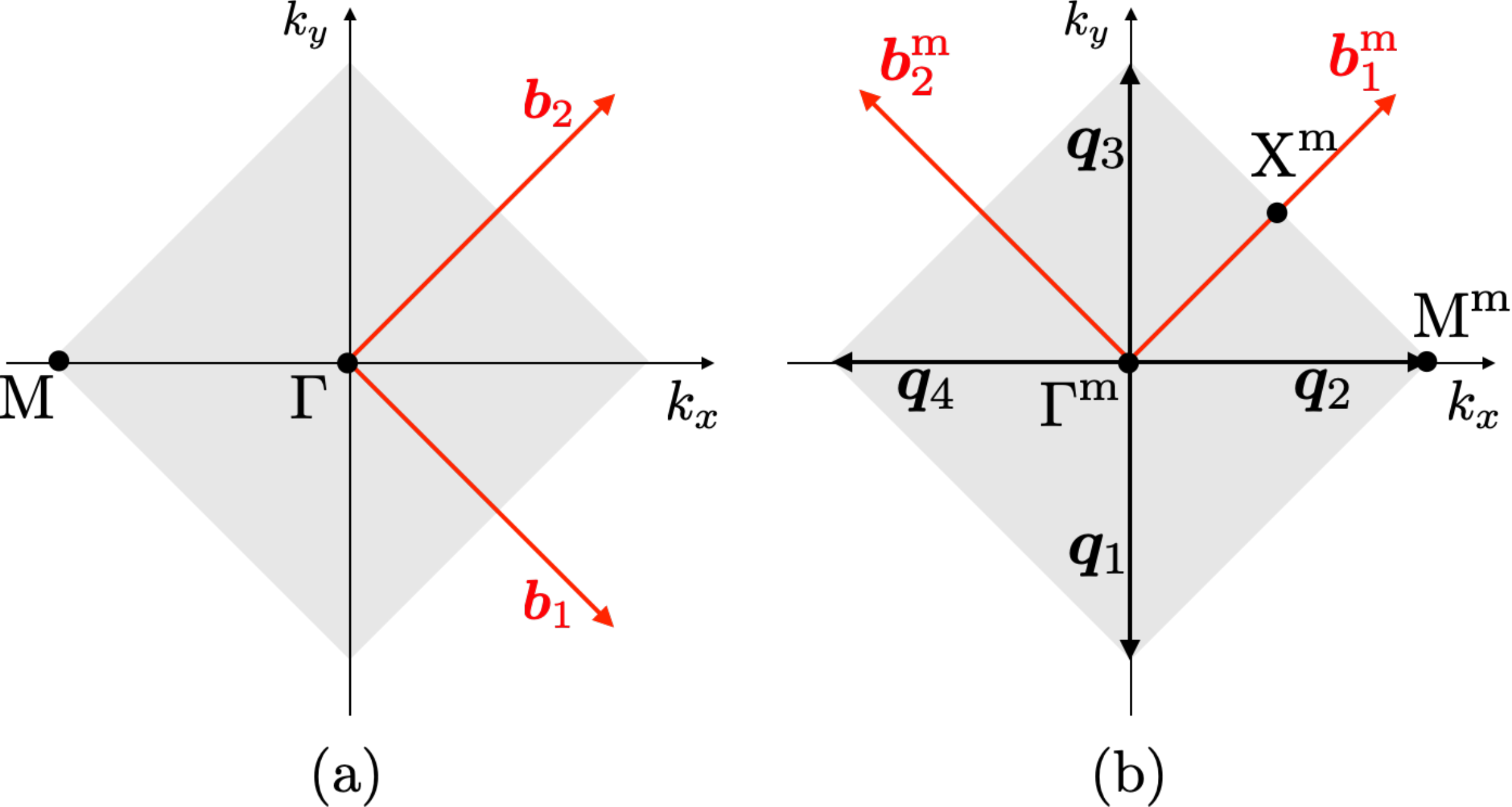}
\end{tabular}
\caption{
(a) Brillouin zone and the primitive reciprocal vectors of the square lattice tight-binding model with $\pi$ flux.  
(b) Mini Brillouin zone and its primitive reciprocal vectors of the moire system.
}
\label{f:brill}
\end{center}
\end{figure}

Next, let us consider the Hamiltonian in the momentum space.
The primitive reciprocal vectors are 
$\bm b_1=(2\pi/a)\left(1/2,-1/2\right)$
 and $\bm b_2=(2\pi/a)\left(1/2,1/2\right)$, as illustrated in Fig. \ref{f:brill} (a).
 The momentum vector is defined by $\bm k=\xi_\mu\bm b_\mu$, where $\xi_\mu=a k_\mu/(2\pi)$ with $\mu=1$ and $2$.
We also define the $xy$ components of $\bm k$ as $ak_x=\bm k\cdot \bm \tau_x$ and $ak_y=\bm k\cdot \bm \tau_y$.
In the momentum representation, 
\begin{alignat}1
c_{A(B),\bm r}=\int_{-\pi/a}^{\pi/a}\frac{d^2k}{(2\pi/a)^2}e^{i\bm k\cdot\bm r}c_{A(B),\bm k},
\end{alignat}
the Hamiltonian becomes
\begin{alignat}1
&H=\int_{-\pi/a}^{\pi/a}\frac{d^2k}{(2\pi/a)^2}\bm c_{\bm k}^\dagger
{\cal H}(k_x,k_y)
\bm c_{\bm k}
\nonumber\\
&{\cal H}(k_x,k_y)\equiv
\left(\begin{array}{cc}\Delta&h(k_x,k_y) \\ \bar h(k_x,k_y)&-\Delta \end{array} \right),
\label{MomHam}
\end{alignat}
where $h(k_x,k_y)=-2t(\sin ak_x+i\sin ak_y)$ and $d^2k\equiv dk_1dk_2=2dk_xdk_y$.
In the rotated frame, let us define $ak'_x=\bm k\cdot\bm \tau_x'$ and $ak'_y=\bm k\cdot\bm \tau_y'$.
It then follows from Eq. (\ref{TraLawTau})
that the transformation law of the wave vector under $\pi/2$ rotation  is
\begin{alignat}1
k_x'=-k_y,\quad k_y'=k_x.
\end{alignat}
Then, the $C_4$ invariance is manifest: 
\begin{alignat}1
{\cal H}(k_x,k_y)={\cal H}(k_y',-k_x')=W {\cal H}(k'_x,k'_y)W^\dagger.
\end{alignat}
It is also easy to see that when $\Delta=0$, the Hamiltonian (\ref{MomHam}) has TR symmetry,
${\cal T}{\cal H}(\bm k){\cal T}^{-1}={\cal H}(-\bm k)$, where ${\cal T}$ is given by ${\cal T}=i\sigma^2{\cal K}$
with the complex conjugation operator ${\cal K}$.

\subsection{Continuum limit}\label{s:continuum}

From Eq. (\ref{MomHam}), one finds  that
the gapless points appear at $\bm k_{\rm \Gamma}=a^{-1}(0,0)$ and $\bm k_{\rm M}=a^{-1}(-\pi,0)$ shown 
in Fig. \ref{f:brill}.
Around $\bm k_{\rm M}$, we have
\begin{alignat}1
&2ta\left(\begin{array}{cc}&k_x-ik_y\\ k_x+ik_y&\end{array}\right)
\nonumber\\
&\rightarrow 2ta\left(\begin{array}{cc}&-2i\partial\\ -2i\bar\partial&\end{array}\right)
=-iv_0\sigma^\mu\partial_\mu,
\label{ConLim}
\end{alignat}
where $v_0=2ta$, the metric tensor is $g_{\mu\nu}={\rm diag}(1,1)$, 
\begin{alignat}1
\partial=\partial_z\equiv \frac{1}{2}(\partial_x-i\partial_y),\quad \bar\partial=\partial_{\bar z}\equiv \frac{1}{2}(\partial_x+i\partial_y),
\end{alignat}
with $z=x+iy$ and $\bar z=x-iy$.
The arrow in Eq. (\ref{ConLim})  represents the switch from momentum space to (continuum) coordinate space.
Thus, together with the staggered potential, total Hamiltonian becomes 
\begin{alignat}1
{\cal H}_{\rm M}(\bm x)&\equiv v_0\left(-i\sigma^\mu\partial_\mu+m\sigma^3\right)
\nonumber\\&
=v_0(-i\bm\sigma\cdot \bm\nabla+m\sigma^3),
\end{alignat}
where $m=\Delta/v_0$.

On the other hand, around $\bm k_{\rm \Gamma}$, we have
\begin{alignat}1
&-2at\left(\begin{array}{cc}&k_x+ik_y\\ k_x-ik_y&\end{array}\right)
\nonumber\\
&\rightarrow -2ta\left(\begin{array}{cc}&-2i\bar\partial\\ -2i\partial&\end{array}\right)
=iv_0\sigma^\mu\partial_\mu,
\end{alignat}
where the metric tensor is $g_{\mu\nu}={\rm diag}(1,-1)$.

\subsection{Transformation law under $SO(2)$ rotation}

The Hamiltonian (\ref{ConLim}) has continuous rotational symmetry.
For the $\theta$ rotation, $\bm \tau'_j=R_{-\theta}\bm \tau_j$,
the coordinate becomes $\bm x'=R_\theta\bm x$, or  
\begin{alignat}1
z'=e^{i\theta}z, \quad \bar z'=e^{-i\theta}\bar z,
\end{alignat}
and hence, 
\begin{alignat}1
\partial'=e^{- i\theta}\partial,\quad \bar\partial'=e^{ i\theta}\bar\partial.
\end{alignat}
Under this, the Hamiltonian transforms as
\begin{alignat}1
{\cal H}_{\rm M}(\bm x)&\equiv v_0\left(\begin{array}{cc}m&-2i\partial\\ -2i\bar\partial&-m\end{array}\right)
\nonumber\\
&=v_0\left(\begin{array}{cc}m&-2ie^{i\theta}\partial'\\ -2ie^{-i\theta}\bar\partial'&-m\end{array}\right)
\nonumber\\
&=W_\theta^\dagger v_0\left(\begin{array}{cc}m&-2i\partial'\\ -2i\bar\partial'&-m\end{array}\right)W_\theta,
\end{alignat}
where $W_\theta={\rm diag}(1,e^{i\theta})$.
Therefore rotational symmetry is denoted by
\begin{alignat}1
\hat{\cal H}_{\rm M}(\bm x)= \hat{\cal H}_{\rm M}(R_{-\theta}\bm x')
= W_\theta^\dagger \hat{\cal H}_{\rm M}(\bm x') W_\theta.
\label{DirXO2}
\end{alignat}
In the special case, $\theta=\pi/2$, 
the transformation law of the coordinates are $x'=-y$ and $y'=x$, i.e., 
$z'=iz$ and $\bar z'=-i\bar z$, and $W_{\pi/2}=W$ in Eq. (\ref{GauTra}).
Thus,  the $\pi/2$ rotational invariance is denoted by
\begin{alignat}1
\hat{\cal H}_{\rm M}(x,y)= \hat{\cal H}_{\rm M}(y',-x')= W^\dagger \hat{\cal H}_{\rm M}(x',y') W.
\label{DirXC4}
\end{alignat}

\section{Twisted bilayer system}\label{s:twist}

Based on the Dirac Hamiltonian obtained in Sec. \ref{s:continuum}, 
we derive the BM Hamiltonian for the twisted bilayer system in this section 
\cite{Bistritzer:2011ab,Moon:2013vv,ledwith2021strong}.
To this end, let us introduce the Dirac momentum $\bm k_{\rm M}$ 
explicitly in the Hamiltonian by replacing 
$-2i\partial\rightarrow -2i\partial-\bar k_{\rm M}$ and
$-2i\bar\partial\rightarrow -2i\bar\partial-k_{\rm M}$, 
where $k=k_x+ik_y$ and $\bar k=k_x-ik_y$ denote the complex momenta. 
Such a Hamiltonian  is denoted by $\widetilde{\cal H}$, where the tilde means that it includes the momenta 
at the M point, but the subscript M for the Hamiltonian is suppressed for simplicity.
%
The twisted system is obtained by stacking the identical but mutually $\pm\theta/2$-rotated systems.
The Hamiltonian of the upper $\theta/2$-rotated system is obtained such that
\begin{alignat}1
\widetilde{\cal H}^{+}&=v_0\left(\begin{array}{cc}m&-2i\partial-\bar k_{\rm M}\\ -2i\bar\partial-k_{\rm M}&-m\end{array}\right)
\nonumber\\
&=v_0\left(\begin{array}{cc}m&e^{ i\theta/2}(-2i\partial')-\bar k_{\rm M}\\ 
e^{- i\theta/2}(-2i\bar\partial')-k_{\rm M}&-m\end{array}\right)
\nonumber\\
&\equiv 
v_0[\bm\sigma_{\theta/2}\cdot(-i\nabla'-\bm k^{+}_{\rm M})+m\sigma^3],
\end{alignat}
where
$\bm\sigma_{\theta/2}=e^{i(\theta/4)\sigma^3}(\sigma^1,\sigma^2)e^{- i(\theta/4)\sigma^3}$ and 
$\bm k^{\pm}_{\rm M}=R_{\pm\theta/2}\bm k_{\rm M}$.
In what follows, the coordinate $\bm x'$ is simply denoted by $\bm x$.  
Together with the lower $-\theta/2$-rotated system $\widetilde{\cal H}^-$, 
the twisted bilayer system  can be described by
\begin{alignat}1
\widetilde{\cal H}&=
\left(\begin{array}{cc}\widetilde{\cal H}^{+}&\widetilde U\\ \widetilde U^\dagger & \widetilde{\cal H}^{-}\end{array}\right)
=U_0^\dagger\left(\begin{array}{cc}{\cal H}_0^{+}&  U\\ \ U^\dagger & {\cal H}_0^{-}
\end{array}\right)U_0
\equiv U_0^\dagger {\cal H}U_0,
\label{TwiHamBas}
\end{alignat}
where 
$U_0={\rm diag}(e^{-i\bm k^+_{\rm M}\cdot\bm x},e^{-i\bm k^-_{\rm M}\cdot\bm x})
$, and
\begin{alignat}1
&{\cal H}_0^\pm=v_0(-i\bm\sigma_{\pm\theta/2}\cdot\nabla+m\sigma^3),
\nonumber\\
&U=\widetilde U e^{-i\bm q_1\cdot\bm x}.
\label{UniTra}
\end{alignat}
Note that the unitary transformation by $U_0$ induces $e^{\pm i\bm q_1\cdot\bm x}$ 
factors for the off-diagonal terms in ${\cal H}$ as noted in Eq. (\ref{UniTra}), where  
$\bm q_1=\bm k^+_{\rm M}-\bm k^-_{\rm M}=k_\theta (0,-1)$ with
$k_\theta=2k_F\sin(\theta/2)$. Here,  $k_F=\pi/a$.
In what follows, we study ${\cal H}$ in Eq. (\ref{TwiHamBas}) as the Hamiltonian for the 
twisted bilayer system.

Although we have the other Dirac Hamiltonian at the $\Gamma$ point, it does not contribute to
the moir\'e interference because of vanishing momentum: it remains a simple massive Dirac fermion even in the bilayer system. 

\subsection{Interlayer coupling: symmetry argument}\label{s:IC_sym}

As discussed in the previous section, there should appear $e^{\pm i\bm q_1\cdot\bm x}$ dependence in 
$U$ introduced in Eq. (\ref{TwiHamBas}).
Therefore, for the interlayer coupling $U(\bm x)$ to be $C_4$-symmetric, it should be of the form
\begin{alignat}1
U(x,y)=\sum_{j=1}^4 U_{j}e^{-i\bm q_j\cdot\bm x},
\label{UPot}
\end{alignat}
where $\bm q_{j}=R_{(j-1)\pi/2}\bm q_1$. 
$C_4$ symmetry (\ref{DirXC4}) requires
\begin{alignat}1
U(x,y)=W^\dagger U(-y,x) W.
\end{alignat}
It then follows that
\begin{alignat}1
\sum_{j=1}^4 U_{j}e^{-i\bm q_{j}\cdot\bm x}=\sum_{j=1}^4  W^\dagger U_{j} W e^{-i\bm q_{j-1}\cdot\bm x}.
\end{alignat}
Thus, 
\begin{alignat}1
U_{j+1}= W U_{j} W^\dagger.
\label{WCon}
\end{alignat}
Let us set $U_{1}=\sum_{j=0}^3w_j\sigma^j$, where $\sigma^0=\1$ and four parameters $w_j$  are complex numbers
in general. 
Then, it follows from Eq. (\ref{WCon}) that generic interlayer coupling (\ref{UPot}) reads
\begin{alignat}1
U&(\bm x)=
\left(
\begin{array}{cc}
w_+v(\bm x)&\bar w\bar u(-\bm x)\\ wu(\bm x)&w_-v(\bm x)
\end{array}
\right),
\label{InterSym}
\end{alignat}
where $w_\pm=w_0+w_3$, $w=w_1+iw_2$, and 
\begin{alignat}1
v(\bm x)
&=\sum_{j=1}^4 e^{-i\bm q_j\cdot\bm x}
=2(\cos k_\theta x+\cos k_\theta  y),
\nonumber\\
u(\bm x)
&=
\sum_{j=1}^4\omega^{j-1}e^{-i\bm q_j\cdot\bm x}
= 2(\sin k_\theta x+i\sin k_\theta y) .
\label{VU}
\end{alignat}
Here, $\omega$ is defined in Eq. (\ref{GauTra}). 
In the following, we assume that the $AA$ and $BB$ couplings
are the same, $w_+=w_-=w_0$.
TR symmetry requires $\bar w_0=w_0$. 
Therefore, the interlayer potentials are generically governed by two parameters, real $w_0$ and complex $w$.

\subsection{Interlayer coupling:  microscopic derivation}\label{s:IC_micro}

So far we have considered the twisted system of the $\pi$-flux model. Even under a uniform external field,
it is very hard to write the concrete interlayer coupling of the twisted lattice model  via explicit gauge-fixing.
For this reason, in the previous section \ref{s:IC_sym}, we have derived the interlayer coupling only on the basis of the symmetry argument.
In this section, we show that there exists a concrete lattice model which yields the interlayer coupling (\ref{InterSym})
in the continuum limit of the BM type.

\subsubsection{Lattice model}\label{s:lattice}

In the incommensurate case, the $x$-$y$ positions of the A and B of the upper and lower layers never match. 
To be concrete, 
let $\bm r^+=\bm r$ and $\bm r^-=\tilde{\bm r}-\bm\tau_zd$ 
be the position vectors for the upper and lower layers, respectively, separated by $d$,
where $\bm r$ and $\tilde{\bm r}$ stand for lattice point vectors on the 2D upper and lower $x$-$y$ planes.
Note that for $\bm r^+-\bm r^-=\bm r-\tilde{\bm r}+\bm \tau_zd$, 2D site positions never match, 
$\bm r-\tilde{\bm r}\ne0$ in the incommensurate case.
Thus, we set $\bm r-\tilde{\bm r}=r(\cos\theta,\sin\theta)$.

The interlayer coupling is generically written as 
\begin{alignat}1
H_{\rm int}&=
\sum_{a,b}\sum_{\bm r^+,\bm r^-}t_{ab}(\bm r^+-\bm r^-)c^\dagger_{a,\bm r^+} \tilde c_{b,\bm r^-}+\rm H.c
\nonumber\\
&\equiv \sum_{a,b}H_{ab}+\rm H.c,
\label{IntLay}
\end{alignat}
where $a,b$ take $A, B$, and $\tilde c_{a,\bm r^-}$ is the annihilation operator of the lower layer. 
We assume 
\begin{alignat}1
&t_{AA}(\bm r^+-\bm r^-)=t_{BB}(\bm r^+-\bm r^-)=f(r),
\nonumber\\
&t_{AB}(\bm r^+-\bm r^-)=f(r)e^{i(\theta+\theta_0)},
\nonumber\\
&t_{BA}(\bm r^+-\bm r^-)=f(r)e^{-i(\theta-\theta_0)},
\label{IntLay2}
\end{alignat}
where the right-hand-sides are parameterized by $\bm r-\tilde{\bm r}=r(\cos\theta,\sin\theta)$ and $\theta_0=\pi/2$.
The $AB$ coupling is the same form as introduced Sec. \ref{s:single} for the single-layer system.
Generically, $f(r)$ is a function of $\sqrt{r^2+z_0^2}$, where $z_0=da$, which is sufficiently short-ranged.
In Appendix \ref{a:fourier}, we discuss a specific form of $f(r)$ as well as its Fourier transformation. 
We have introduced the interlayer coupling in Eq. (\ref{IntLay2}) because it keeps the key symmetries of the 
$\pi$-flux model, $C_4$ symmetry and TR symmetry:
for the $\pi/2$ rotation, from the transformation law, $(x-\tilde x,y-\tilde y)=(y'-\tilde y',-(x'-\tilde x'))$, implying that
$\theta=\theta'-\pi/2$, as well as from the gauge transformation $c_{A,\bm r^\pm}=c'_{A,\bm r^\pm}$ 
and $c_{B,\bm r^\pm}=ic'_{B,\bm r^\pm}$,
it turns out that the above interlayer coupling  is manifestly $C_4$-invariant.
It is also TR invariant, since the relations $t_{AB}^*(\bm r^+-\bm r^-)=-t_{BA}(\bm r^+-\bm r^-)$ holds.

\subsubsection{Low-energy interlayer coupling}\label{s:interlayer}

When a specific lattice model is given, we can derive the effective interlayer coupling, according to BM \cite{Bistritzer:2011ab}.
As summarized in Appendix \ref{a:interlayer}, the microscopic interlayer coupling $H_{ab}$ in Eq. (\ref{IntLay}) 
becomes the following effective potential for the continuum Dirac fermions,
\begin{alignat}1
H_{ab}
=\int_{-\infty}^\infty \frac{d^2x}{a^2}c_{a,\bm x}^\dagger\widetilde U_{ab}(\bm x)\tilde c_{b,\bm x},
\label{IntCouGen1}
\end{alignat}
where 
\begin{alignat}1
\widetilde U_{ab}(\bm x)=\sum_{\bm G,\tilde{\bm G}}
t_{ab}(\bm k_{\rm M}+\bm G)e^{i\bm G\cdot\bm \tau_a}e^{-i\tilde{\bm G}\cdot\tilde{\bm\tau}_b}
e^{-i(\bm G-\tilde{\bm G})\cdot\bm x}.
\label{IntCouGen2}
\end{alignat}
Here, $t_{ab}(\bm q)$ is the Fourier transformation of the $t_{ab}(\bm r^+-\bm r^-)$ given 
in Eqs. (\ref{IntLay}) and (\ref{IntLay2}), 
and $\bm G$ an $\tilde{\bm G}$ are reciprocal vectors of the upper and lower layers, respectively.

Since the Fourier transformation of $f(r)$ and $f(r)e^{i\theta}$ in Eq. (\ref{IntLay2})
decrease  rapidly as functions of $q=|\bm q|$, as discussed in Appendix \ref{a:fourier}, 
the summations over $\bm G$ and $\tilde{\bm G}$ are  restricted 
within the first Brillouin zone.
Note that $\bm\tau_A=0$ and $\bm \tau_B=\bm \tau_x$, so that $\bm b_j\cdot\bm \tau_B=\pi$.
Thus, we have
\begin{alignat}1
\widetilde U_{AA}(\bm x)\sim& t_{AA}(\bm k_{\rm M})+t_{AA}(\bm k_{\rm M}+\bm b_1)e^{-i\bm b^{\rm m}_1\cdot\bm x}
\nonumber\\
&+t_{AA}(\bm k_{\rm M}+\bm b_2)e^{-i\bm b^{\rm m}_2\cdot\bm x}
\nonumber\\
&+t_{AA}(\bm k_{\rm M}+\bm b_1+\bm b_2)e^{-i(\bm b^{\rm m}_1+\bm b^{\rm m}_2)\cdot\bm x},
\nonumber\\
\widetilde U_{AB}(\bm x)\sim& t_{AB}(\bm k_{\rm M})-t_{AB}(\bm k_{\rm M}+\bm b_1)e^{-i\bm b^{\rm m}_1\cdot\bm x}
\nonumber\\
&-t_{AB}(\bm k_{\rm M}+\bm b_2)e^{-i\bm b^{\rm m}_2\cdot\bm x}
\nonumber\\
&+t_{AB}(\bm k_{\rm M}+\bm b_1+\bm b_2)e^{-i(\bm b^{\rm m}_1+\bm b^{\rm m}_2)\cdot\bm x},
\end{alignat}
where moir\'e reciprocal vectors are defined by 
\begin{alignat}1
\bm b_j^{\rm m}&=\bm b_j^+-\bm b_j^-=(R_{\theta/2}-R_{-\theta/2})\bm b_j
\nonumber\\
&=\left\{
\begin{array}{ll}
\bm q_2-\bm q_1=k_\theta(1,1) \quad& (j=1)\\\bm q_4-\bm q_1=k_\theta(-1,1) \quad& (j=2)
\end{array}\right.,
\label{MoiRec}
\end{alignat}
and $\bm q_j$ in the above are defined in Eq. (\ref{UPot}).
These are illustrated in Fig. \ref{f:brill} (b).
Let us set the Fourier transformations in Eq. (\ref{FouTra}),  
$F_0(k_{\rm M})=w_0$ and $F_1(k_{\rm M})=-w$, where the minus sign 
of the latter is for notational convenience only.
It should be noted that {\it $w$ is a real parameter} for the lattice model introduced in this section.
Therefore the interlayer coupling parameters $w_0$ and $w$ are regarded as real constants in the following discussions.
It follows from Eq. (\ref{FouTra}) that each coefficient reads
\begin{alignat}1
t_{AA}(\bm k_{\rm M})&=t_{AA}(\bm k_{\rm M}+\bm b_1)=t_{AA}(\bm k_{\rm M}+\bm b_2)
\nonumber\\
&=t_{AA}(\bm k_{\rm M}+\bm b_1+\bm b_2)=w_0,
\end{alignat}
and 
\begin{alignat}1
&t_{AB}(\bm k_{\rm M})=w,\, t_{AB}(\bm k_{\rm M}+\bm b_1)=iw,
\nonumber\\
&t_{AB}(\bm k_{\rm M}+\bm b_2)=-iw,\,
t_{AB}(\bm k_{\rm M}+\bm b_1+\bm b_2)=-w.
\end{alignat}
It then turns out that the interlayer coupling $U=\widetilde Ue^{-i\bm q_1\cdot \bm x}$ 
in Eq. (\ref{UniTra}) becomes Eq. (\ref{InterSym}) with a real $w$.
Thus we have shown that there exists an explicit model which has the interlayer coupling (\ref{InterSym}).

The moir\'e reciprocal vectors introduced in Eq. (\ref{MoiRec}) enable us to define moir\'e translational vectors.
Based on them, let us finally mention translation symmetry. 
From Eq. (\ref{MoiRec}), we can define 
the primitive moir\'e translation vectors as
\begin{alignat}1
&\bm a_1^{\rm m}=\frac{\pi}{k_\theta}(1,1),\quad \bm a_2^{\rm m}=\frac{\pi}{k_\theta}(-1,1),
\end{alignat}
which satisfy $\bm a_i^{\rm m}\cdot\bm b_j^{\rm m}=2\pi\delta_{i,j}$.
These are illustrated in Fig. \ref{f:moire}.
From  $\bm q_i\cdot\bm a_j^{\rm m}=\pm\pi$,
valid for any $i$ and $j$, it follows that
\begin{alignat}1
U(\bm x+\bm a_j^{\rm m})=-U(\bm x).
\end{alignat}
Thus, translational symmetry reads
\begin{alignat}1
{\cal H}(\bm x+\bm a_j^{\rm m})=\tau^3{\cal H}(\bm x)\tau^3,
\end{alignat}
where $\tau^3$ acts on the space spanned by the upper and lower layers.

\subsection{Symmetries}\label{s:symmetry}

Before considering symmetry properties of the model, let us fix the representation of the Hamiltonian.
In the effective Dirac Hamiltonian (\ref{TwiHamBas}),
the twist angle $\theta$ is embedded in 
each layer Hamiltonian ${\cal H}_0^\pm$ as well as the interlayer coupling through $\bm q_j$. 
It may be convenient to collect such $\theta$ dependence into the parameters $w_0$ and $w$ \cite{Bistritzer:2011ab,Tarnopolsky:2019aa}. 
To this end, let us make the scale transformation, $\bm x\rightarrow \bm x/k_\theta $ and $m\rightarrow k_\theta m$, 
and the rotation induced by $U_\theta\equiv{\rm diag }(e^{i(\theta/4)\sigma^3},e^{-i(\theta/4)\sigma^3})$.
Then, the Hamiltonian becomes
\begin{alignat}1
{\cal H}(\bm x)&\rightarrow
U_\theta^\dagger {\cal H}U_\theta
\nonumber\\
&=
v_0k_\theta\left(\begin{array}{cc}-i\bm\sigma\cdot\nabla+m\sigma^3& U(\bm x)\\ 
U^\dagger(\bm x)&-i\bm\sigma\cdot\nabla+m\sigma^3\end{array}\right),
\label{Ham_re}
\end{alignat}
where rescaled mass parameter is $m=\Delta/(v_0k_\theta)$, and the rotated and rescaled moir\'e potential $U$ is 
\begin{alignat}1
U(\bm x)&\rightarrow \frac{1}{v_0k_\theta}
e^{-i(\theta/4)\sigma^3}U(\bm x)e^{-i(\theta/4)\sigma^3}
\nonumber\\
&=\left(\begin{array}{cc}\alpha_0e^{-i\theta/2}v(\bm x)&-\alpha\bar u(\bm x)\\ 
\alpha u(\bm x)&\alpha_0 e^{i\theta/2} v(\bm x)\end{array}\right),
\label{FinHam}
\end{alignat}
with
$\alpha_0=w_j/(v_0k_\theta)$  and $\alpha=w/(v_0k_\theta)$. 
Note here that $u(\bm x)$ and $v(\bm x)$ are defined by 
Eq. (\ref{VU}) but setting $k_\theta=1$ by rescaling, and also that $u(\bm x)$ is an odd function of $\bm x$.
One finds the explicit expression of the rotated and rescaled Hamiltonian (\ref{Ham_re}) with (\ref{FinHam})
in Eq. (\ref{FinHamA}).

Let us now discuss the symmetries of the above Hamiltonian 
besides $C_4$ symmetry.
In the massless case, $m=0$, the model has TR symmetry 
\begin{alignat}1
&{\cal T} {\cal H}(\bm x){\cal T}^{-1}={\cal H}(\bm x), \quad {\cal T}=i\sigma^2\tau^0{\cal K},
\label{TRS}
\end{alignat}
where ${\cal T}$ operator has already been defined in Sec. \ref{s:lattice_single}.
TR symmetry is broken solely by the mass term originated from 
the staggered potential $\Delta$ in the lattice model.
Therefore this model describes the moir\'e system {\it without an external magnetic field except for $\pi$ flux.}
Also in the massless case, the model has ${\cal C}_{2z}{\cal K}$ symmetry denoted by
\begin{alignat}1
&{\cal C}_{2z}{\cal  K}{\cal H}(\bm x)({\cal C}_{2z}{\cal K})^{-1}={\cal H}(\bm x), \quad 
{\cal C}_{2z}{\cal K}=\sigma^1\tau^0{\cal K},
\label{C2zT}
\end{alignat}
which causes the fragile topology in the TBG. Here, ${\cal C}_{2z}$ denotes the two-fold rotation around the $z$-axis.
When, $\alpha_0=0$ as well as $m=0$, the model has chiral symmetry
\begin{alignat}1
&\Gamma{\cal H}(\bm x)\Gamma^{-1}=-{\cal H}({\bm x}), \quad \Gamma=\sigma^3\tau^0,
\end{alignat} 
which is responsible for the flat bands in the TBG \cite{Tarnopolsky:2019aa}.
In the massive case, $m\ne0$, but $\alpha_0=0$, the model has inversion symmetry
\begin{alignat}1
{\cal P}{\cal H}(\bm x){\cal P}^{-1}={\cal H}(-{\bm x}), \quad {\cal P}=\sigma^3i\tau^2.
\end{alignat} 

In what follows, we consider mainly the massive model with generic interlayer couplings $\alpha_0$ and  
$\alpha_1$,  so that the model has no specific symmetries other than fourfold rotational symmetry.

\subsection{Moir\'e band structure}\label{s:numerical}

\begin{figure}[h]
\begin{center}
\begin{tabular}{cc}
\includegraphics[width=0.5\linewidth]{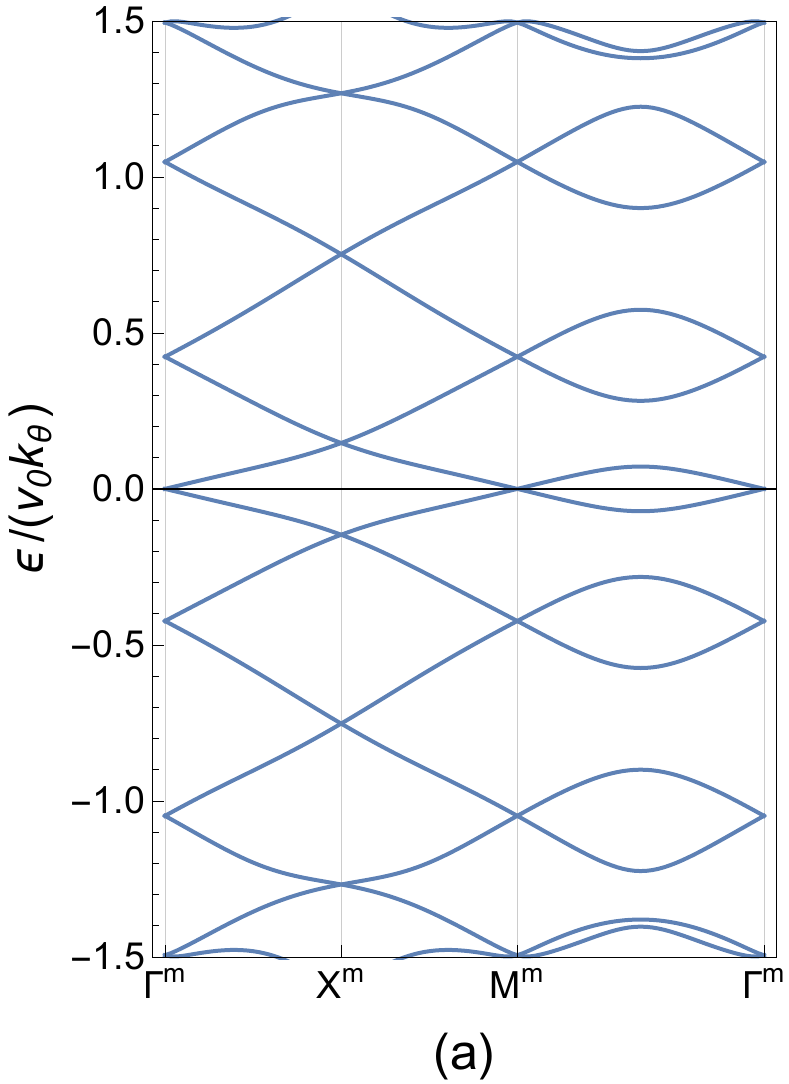}
&
\includegraphics[width=0.5\linewidth]{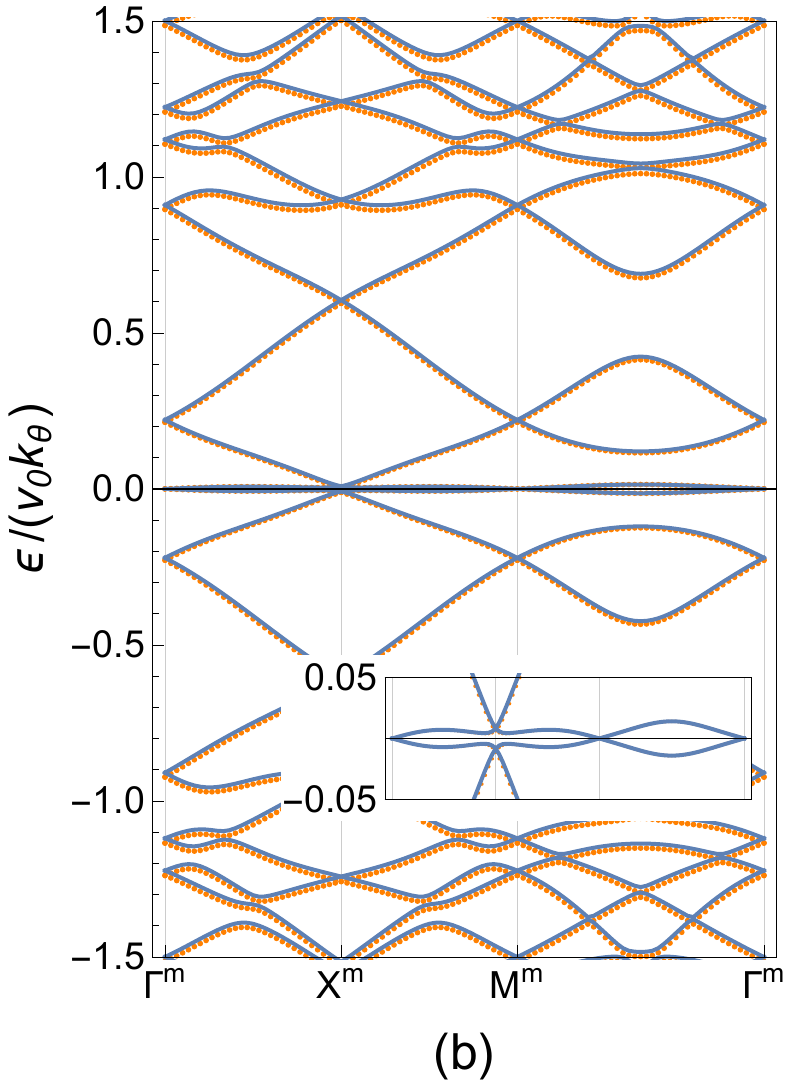}
\end{tabular}
\caption{Spectra of the massless model,
$m=0$. Solid curves are for $\theta=0$, whereas orange dots in (b) are for $\theta=1^\circ$.
The zero energy is set at the band center.
(a) $\alpha_0=0$ and $\alpha=0.5$. (b) $\alpha_0=\alpha=0.5$. Inset shows the spectrum near the 
band center. High-symmetry points on the moir\'e Brillouin zone are defined in Fig. \ref{f:brill} (b).
}
\label{f:sp_m0}
\end{center}
\end{figure}

Now, let us show several characteristic features of the spectrum of the present system.
The $\theta$-dependence of the Hamiltonian (\ref{Ham_re}) is through the renormalized parameters 
$\alpha_0,\alpha=w_0,w/(v_0k_\theta)$ as well as the overall factor $v_0k_\theta$. The exceptional explicit dependence is
only the diagonal stacking potentials $\alpha_0e^{(\pm i\theta/2)}v(\bm x)$ in Eq. (\ref{FinHam}).
Thus we regard $\alpha_0,\alpha$ and $\theta$ as independent parameters in the following calculations.
Namely, $\theta=0$ means that we set so only in Eq. (\ref{FinHam}) while we keep $\alpha_0$ and $\alpha$ finite.
As the energy is measured in unit of $v_0k_\theta$, the model is therefore characterized by four parameters: 
$\alpha_0$, $\alpha$, $\theta$,  and the mass $m$. 
For the interlayer coupling (\ref{IntCou}), we find that generically $\alpha_0<\alpha$ holds, as seen in Fig. \ref{f:fourier}.
However, the qualitative feature does not depends on the difference between $\alpha_0$ and $\alpha$, so that
we often study the case of $\alpha_0=\alpha$, for simplicity.

\begin{figure}[h]
\begin{center}
\begin{tabular}{cc}
\includegraphics[width=0.5\linewidth]{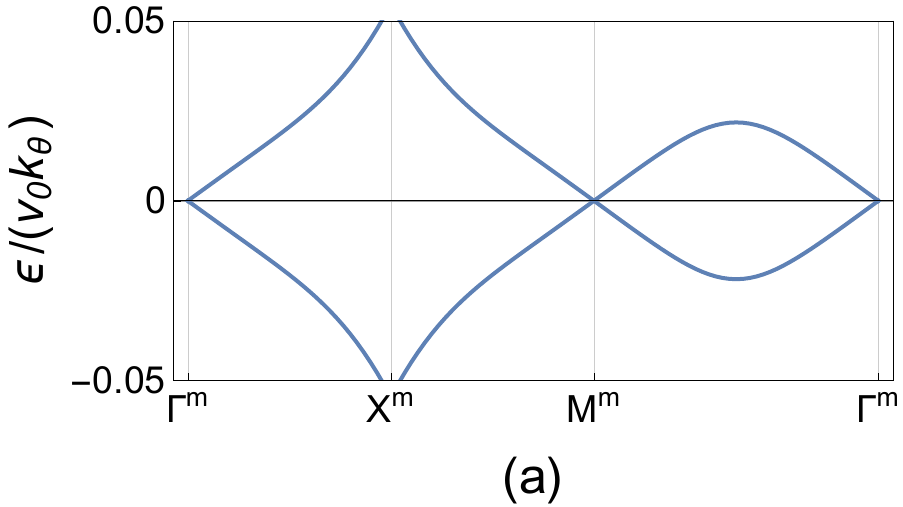}
&
\includegraphics[width=0.5\linewidth]{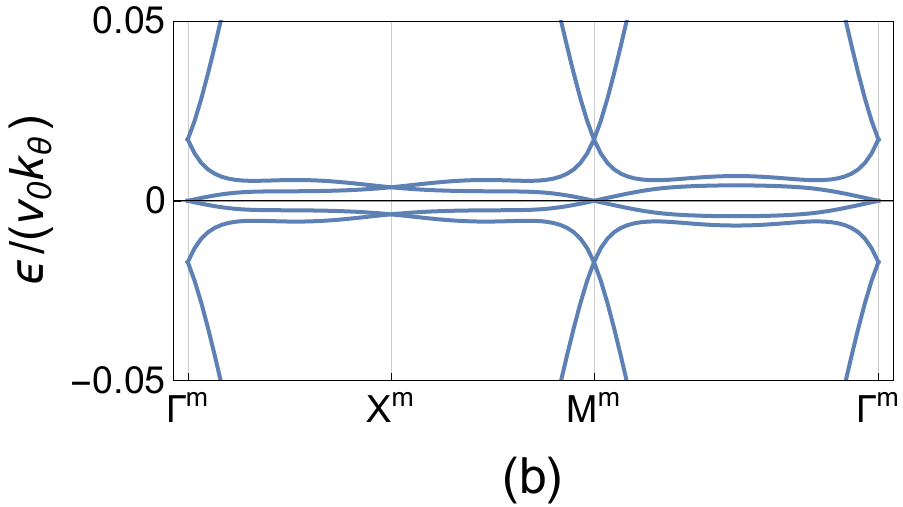}
\\
\includegraphics[width=0.5\linewidth]{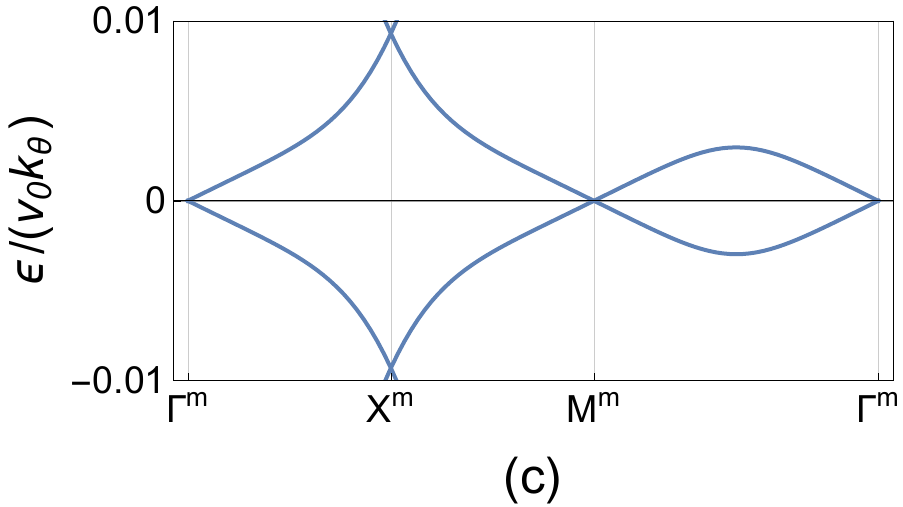}
&
\includegraphics[width=0.5\linewidth]{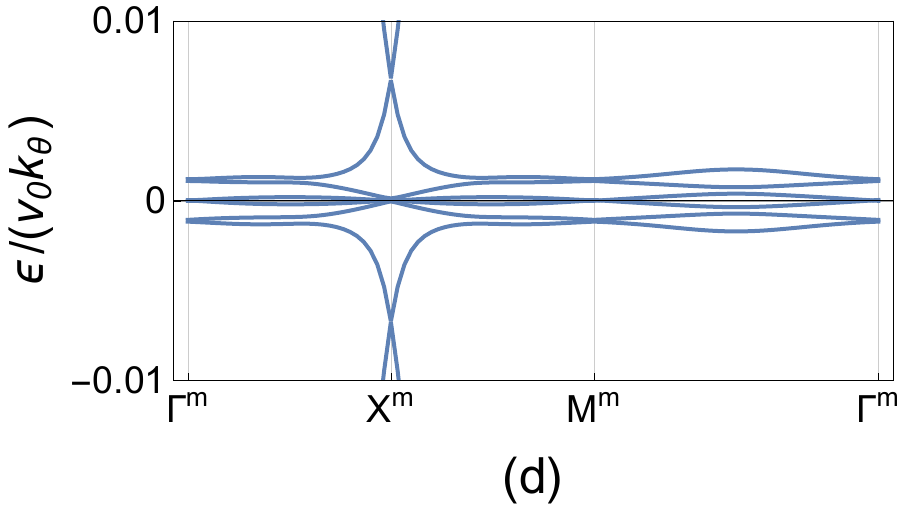}
\end{tabular}
\caption{
Spectra of massless model near the band center for comparison with the inset in Fig. \ref{f:sp_m0} (b).
(a) $\alpha_0=0$ and $\alpha=0.7$, (b) $\alpha_0=\alpha=0.7$,
(c) $\alpha_0=0$ and $\alpha=1.0$, and (d) $\alpha_0=\alpha=1.0$.
}
\label{f:sp_e}
\end{center}
\end{figure}

First, let us consider the massless case where $m=0$.
In Fig. \ref{f:sp_m0} (a), we show the spectrum with a specific parameter $\alpha_0=0$, 
in which the model has chiral symmetry. In the TBG, such a chiral model shows completely flat bands at magic angles,
whose exact wave functions can be obtained \cite{Tarnopolsky:2019aa}. 
The present model, however, does not have any magic angles (special $\alpha$) showing flat bands: 
This is due to TR  symmetry (\ref{TRS}), 
which ensures the Kramers degeneracies at the time-reversal invariant momenta (TRIM), especially at X$^{\rm m}$.
When the potential proportional to $\alpha_0$ is added, which breaks 
the chiral symmetry  but keeps TR symmetry, 
the bands around zero energy become narrower while the overall profiles remain unchanged, 
as shown in Fig. \ref{f:sp_m0} (b).
In this figure, solid-curves and orange dots are calculated for $\theta=0^\circ$ and $\theta=1^\circ$, respectively.
The results tell that the flat band near zero energy is almost independent of $\theta$.
Such a feature is quite generic in the present model. To see this, we compute the spectrum for various parameters
$\alpha$ and $\alpha_0$ in Fig. \ref{f:sp_e}.
Comparing the $\alpha_0=0$ and $\alpha_0\ne0$ cases, 
this figure indicates the importance 
of the $\alpha_0$ potential associated with the $AA$ stacking interlayer coupling:
the flat bands become flatter and more degenerate, as the parameters $\alpha=\alpha_0$ are increased.

\begin{figure}[h]
\begin{center}
\begin{tabular}{cc}
\includegraphics[width=0.5\linewidth]{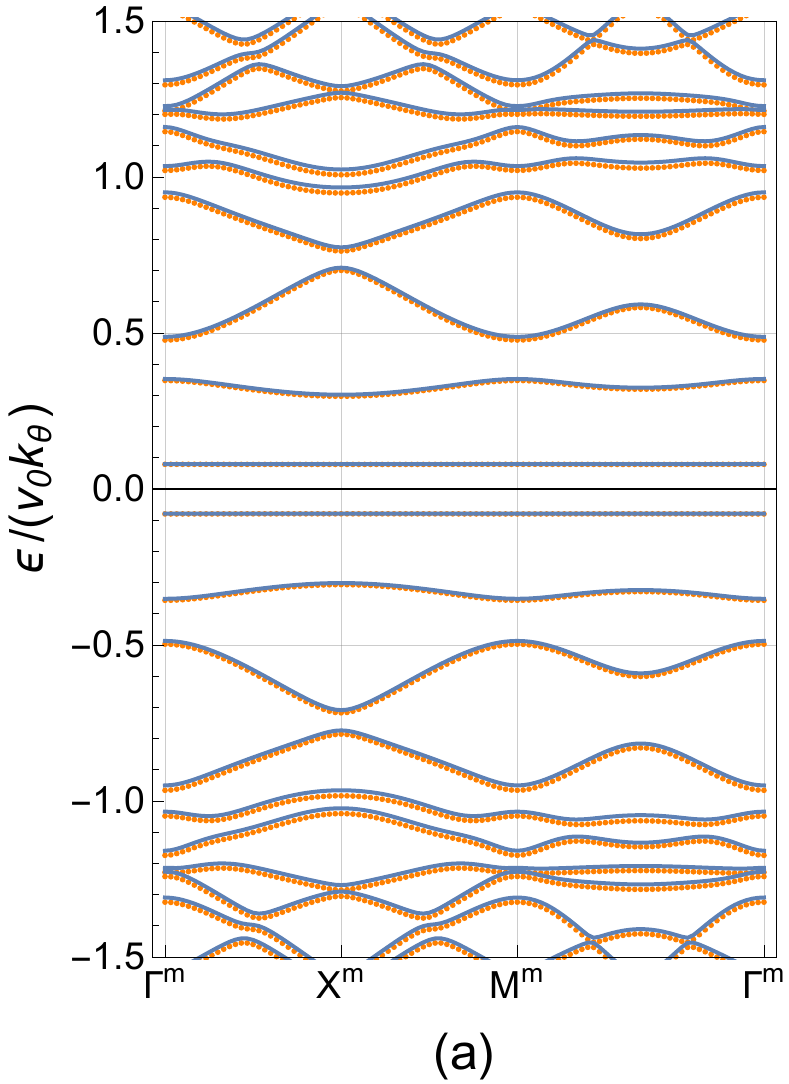}
&
\includegraphics[width=0.5\linewidth]{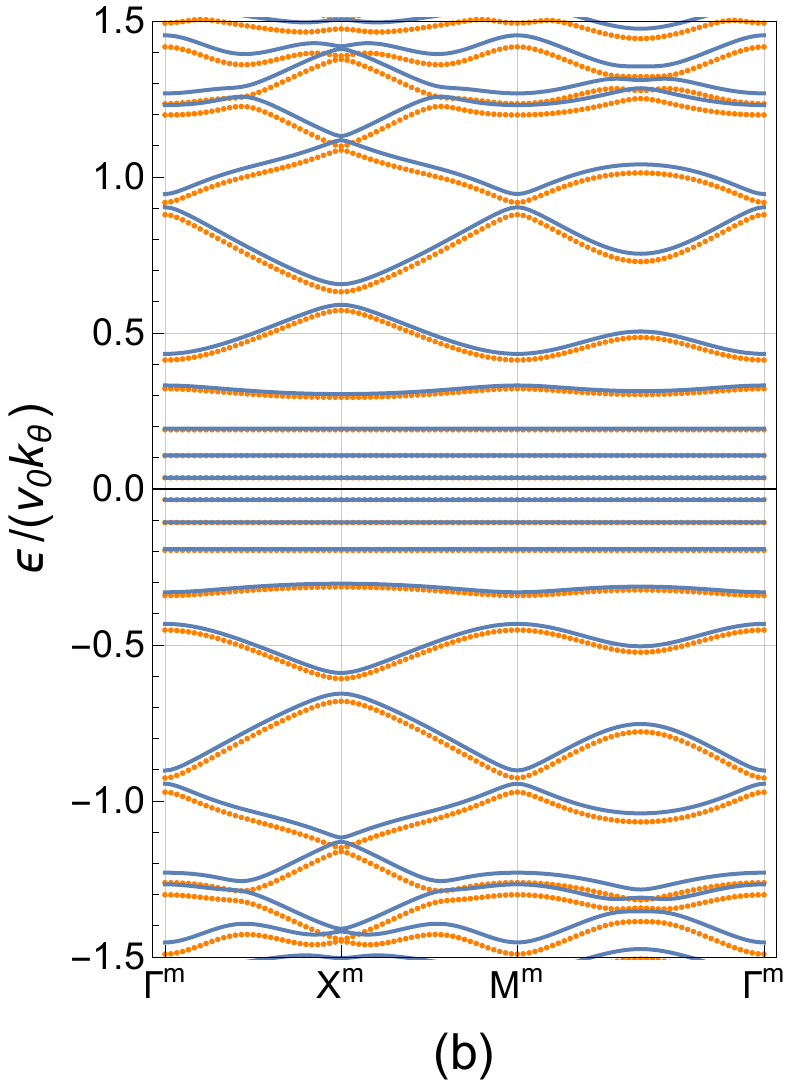}
\end{tabular}
\caption{The spectra of the massive model, $m=0.5$.
(a) $\alpha=\alpha_0=0.5$. (b) $\alpha=\alpha_0=1.0$.
The solid curves and the orange dots are the same as those in Fig. \ref{f:sp_m0}.
}
\label{f:sp_m}
\end{center}
\end{figure}

Next, let us switch to the massive model. 
The present model is quite characteristic in the massive case rather than the massless case, since 
the degeneracy of the flat bands is lifted by the mass term; 
the flat bands of the massless model are separated into 
{\it isolated flat bands}  for the massive model. Moreover, as the parameters $\alpha=\alpha_0$ and/or $m$ are increased,
these flat bands approach even flatter, as shown below.
In Fig. \ref{f:sp_m} (a), we show the spectrum of the massive model with a relatively small parameter
$\alpha(= \alpha_0)$. We see two flat bands around the band center.
Since the solid lines and orange dots coincide, the flat bands at the band center do not depend on $\theta$, 
as in the massless case.

\begin{figure}[h]
\begin{center}
\begin{tabular}{cc}
\includegraphics[width=0.5\linewidth]{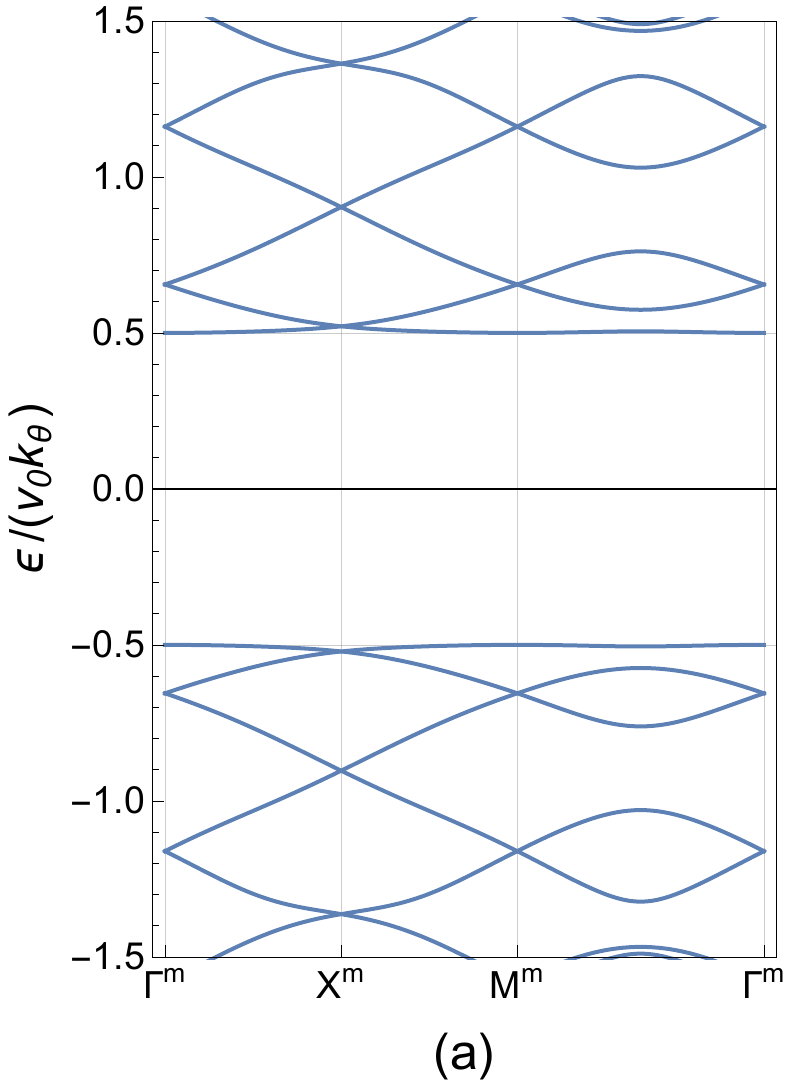}
&
\includegraphics[width=0.5\linewidth]{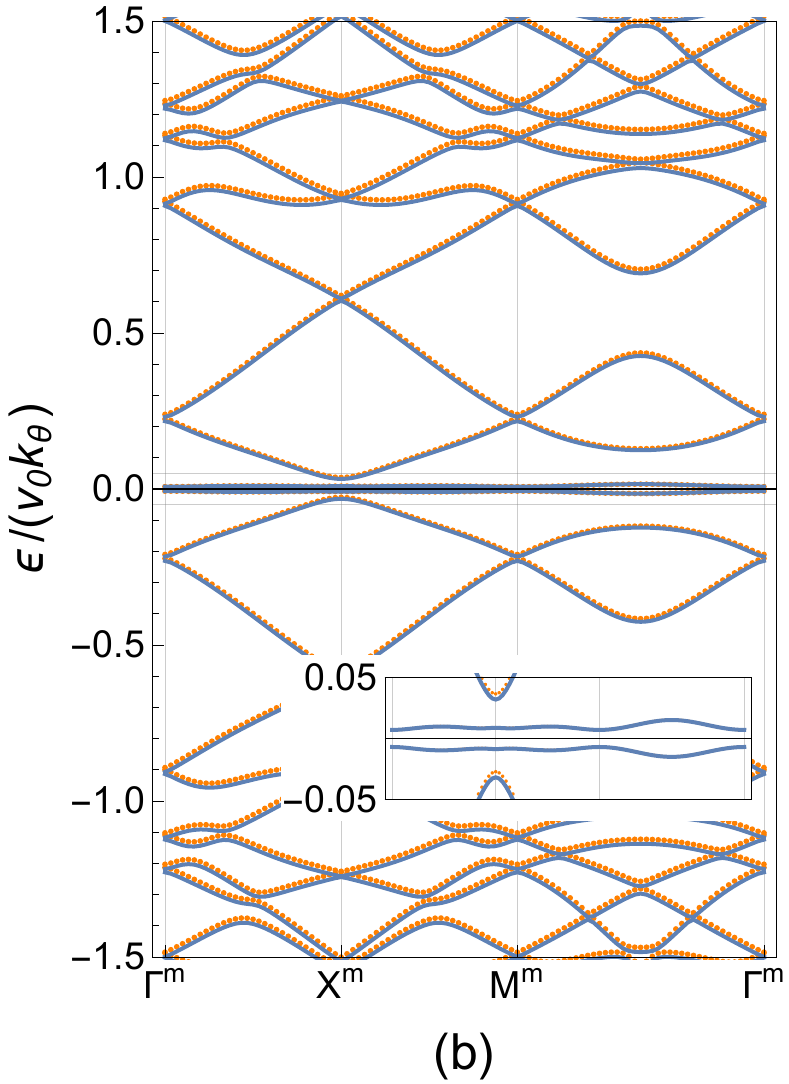}
\end{tabular}
\caption{
Spectra of the massive model.  
(a)  includes a mass $m=0.5$ in Fig. \ref{f:sp_m0} (a), and 
(b) includes a small mass $m=0.05$ in Fig. \ref{f:sp_m0} (b).
The solid curves and the orange dots are the same as those in Fig. \ref{f:sp_m0}.
}
\label{f:sp_a}
\end{center}
\end{figure}

To clarify the nature of the flat band, let us consider separately cases where the parameters are chosen artificially.
Among the three parameters, $\alpha_0$, $\alpha$ and $m$,  other than $\theta$,
we first consider the role of the mass parameter $m$.
To this end, 
let us start with Fig. \ref{f:sp_m0} (b), i.e., $m=0$ but $\alpha=\alpha_0>0$,  and give a small  but finite mass.
As argued above, the Kramers degeneracies at TRIM, especially at X$^{\rm m}$,
are lifted due to the mass term which 
breaks TR symmetry. Then, we see in Fig. \ref{f:sp_a} (b) that two almost flat bands remain around the band center,
separated from others.

\begin{table}[htb]
\caption{
Energies and band widths of the negative energy flat bands in Fig. \ref{f:sp_m} (b). 
These values are in unit of $v_0k_\theta$.  The energies denoted by ``LL energy" are 
computed by the Hamiltonian (\ref{LocHam}).
 }
 \label{t:width}
\begin{ruledtabular}
\begin{tabular}{c|cc|c}
band \# & energy & width & LL energy\\
\hline
1&$-0.035$ & $1.3\times10^{-5} $& $-0.040$\\
2&$-0.107$ & $5.9\times10^{-5} $& $-0.106$\\
3&$-0.193$ & $7.7\times10^{-5} $& $-0.221$\\
4&$-0.331$ & $2.8\times10^{-2} $& $-0.331$\\
\end{tabular}
\end{ruledtabular}
\end{table}

On the other hand, if the parameter $\alpha=\alpha_0=0$, 
the Hamiltonian is a simple  massive Dirac model, allowing no states in between $-m$ and $m$. 
This is also true if only the potential $\alpha$ is included, which will be discussed in Sec. \ref{s:landau}.
Therefore the potential $\alpha_0$ plays a role of yielding states within the mass gap.  
To see this,  we start from Fig. \ref{f:sp_m0} (a), i.e., the case of $\alpha_0=0$ and add the mass $m=0.5$ to it, which 
corresponds to the model in Fig. \ref{f:sp_m} (a) but with $\alpha_0=0$. 
The spectrum of such a model is shown in Fig. \ref{f:sp_a} (a). One can see no states within the mass gap. 
As $\alpha_0$ is increased, some of the bands move into the mass gap and we finally reach Fig. \ref{f:sp_m} (a).

So far we have shown that the massive Dirac Hamiltonian with moir\'e potential allows flat bands, 
which appear within the mass gap. 
The number of such flat bands depends on $\alpha_0$ and $\alpha$: 
Increasing the value of these parameters not only increases the number of flat bands, but also makes them flatter.
In Fig. \ref{f:sp_m} (b), we show the spectrum in the case of larger $\alpha=\alpha_0$. 
There are four flat bands in the negative and positive energies, respectively. 
Their band-widths as well as energies are listed in Table \ref{t:width}.
Very narrow widths are reminiscent of Landau levels under a uniform magnetic field. 
However, not only the lattice model but also the effective Dirac model have TR symmetry when $m=0$. 
Here, $m$ is originated from the staggered potential  on the lattice. 
Therefore, the Hamiltonian does not includes a magnetic field except for $\pi$ flux per plaquette, and
hence, the flat bands cannot be Landau levels induced by an external magnetic field.

\begin{figure*}[htb]
\begin{center}
\begin{tabular}{ccccc}
\includegraphics[width=0.19\linewidth]{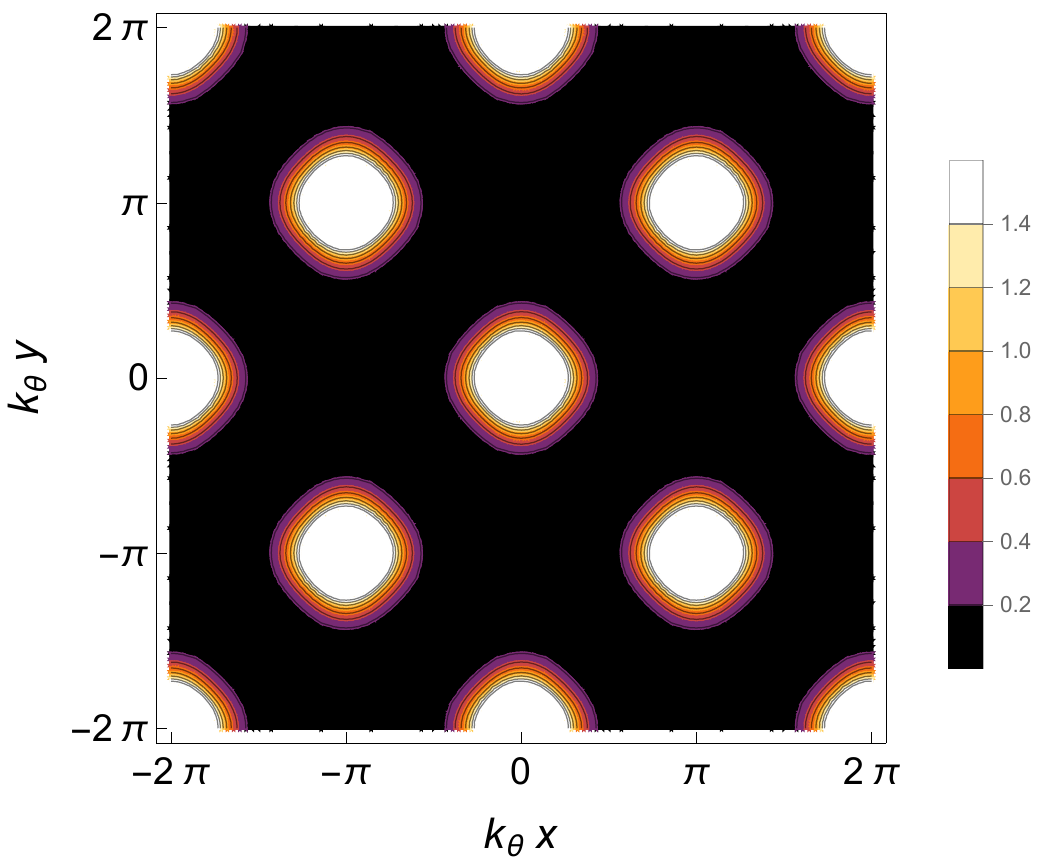}
&
\includegraphics[width=0.19\linewidth]{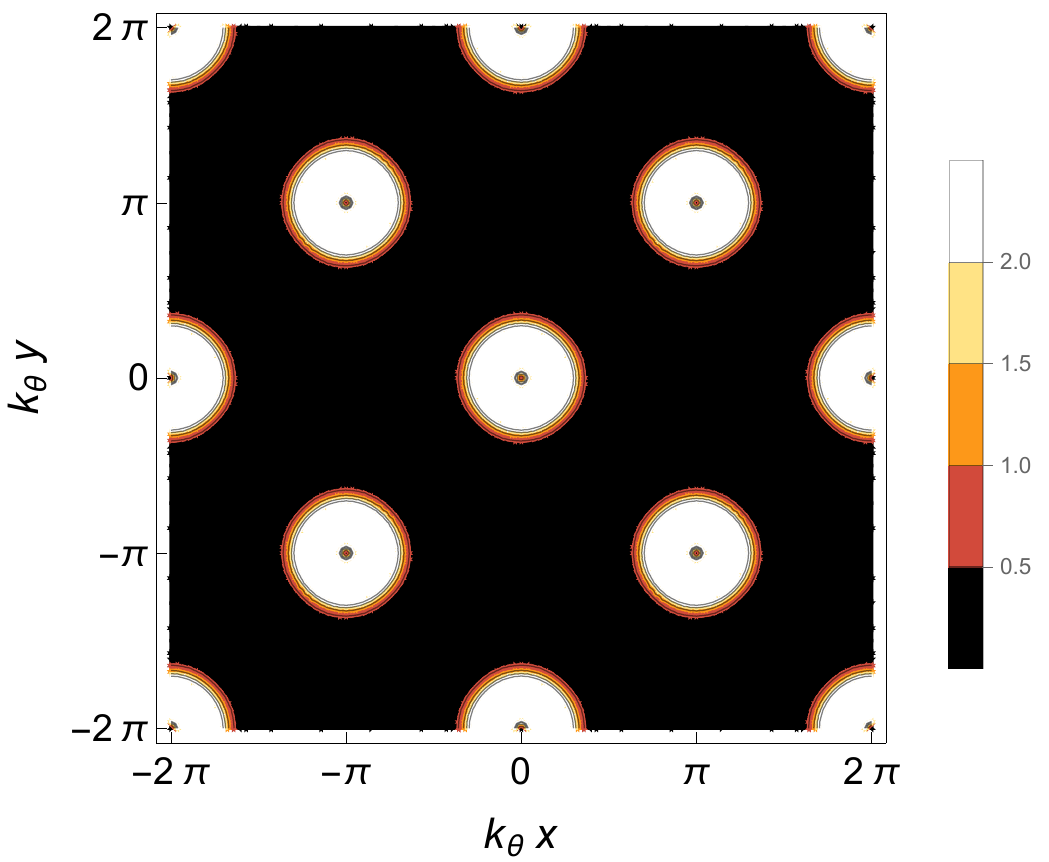}
&
\includegraphics[width=0.19\linewidth]{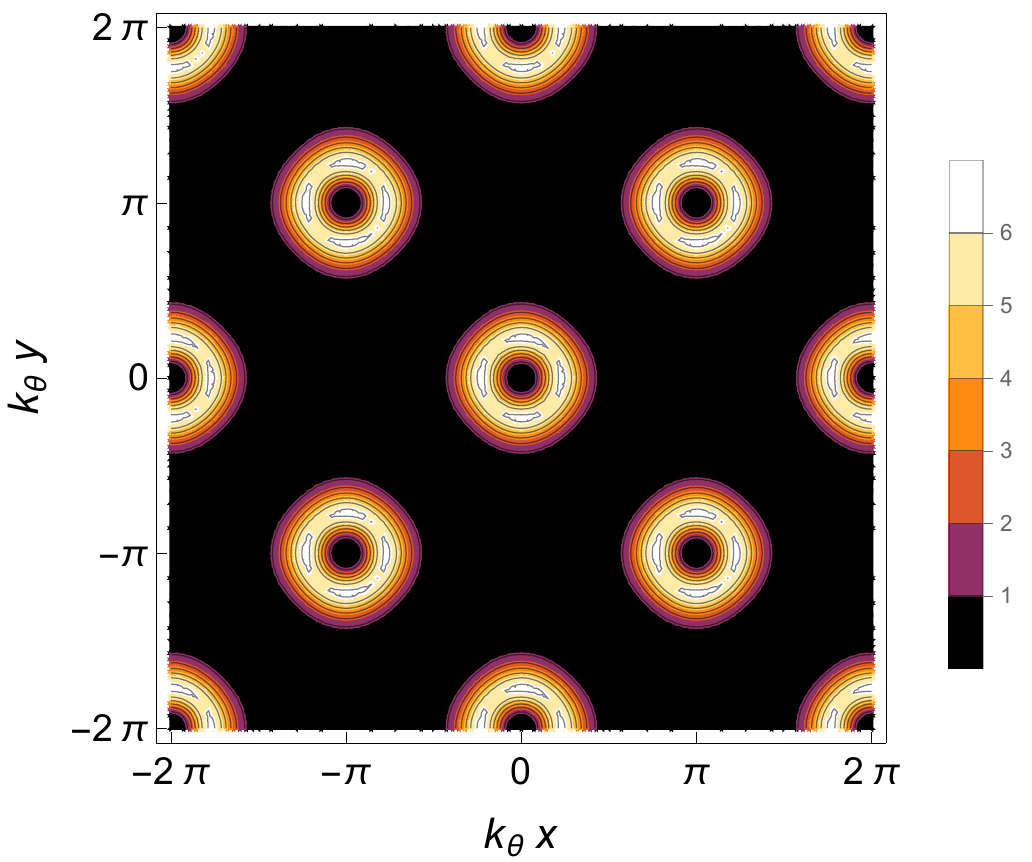}
&
\includegraphics[width=0.19\linewidth]{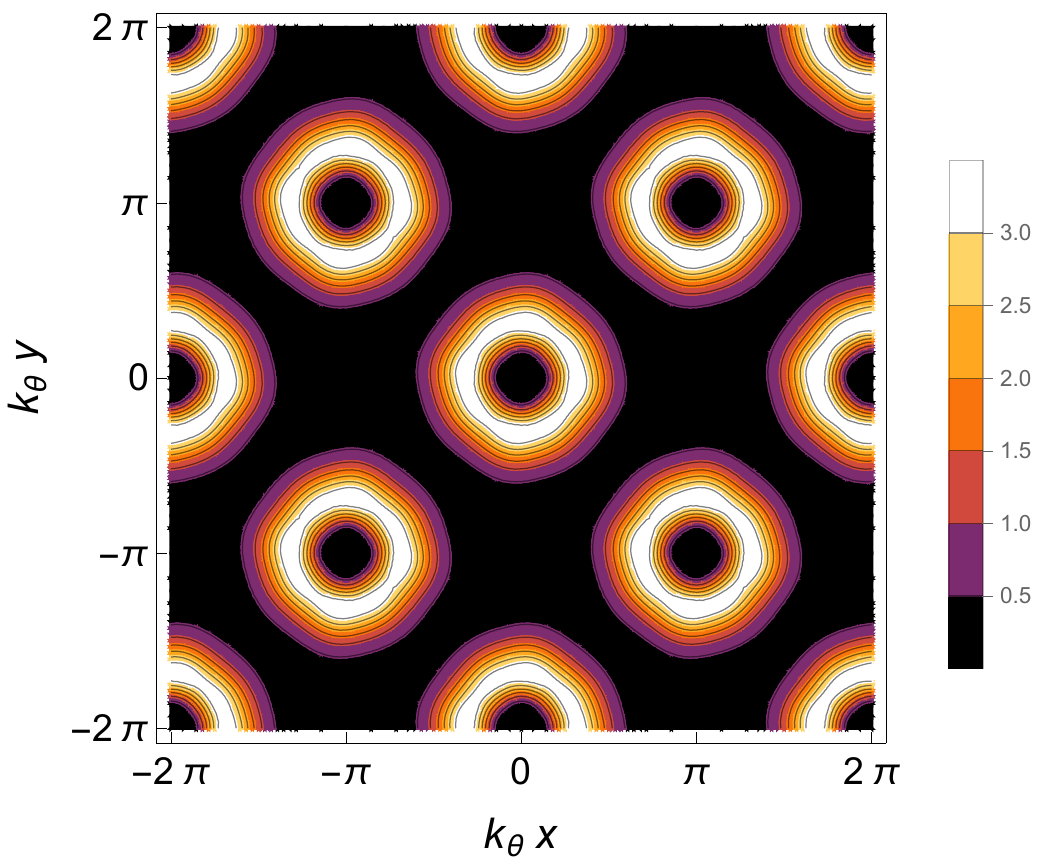}
&
\includegraphics[width=0.19\linewidth]{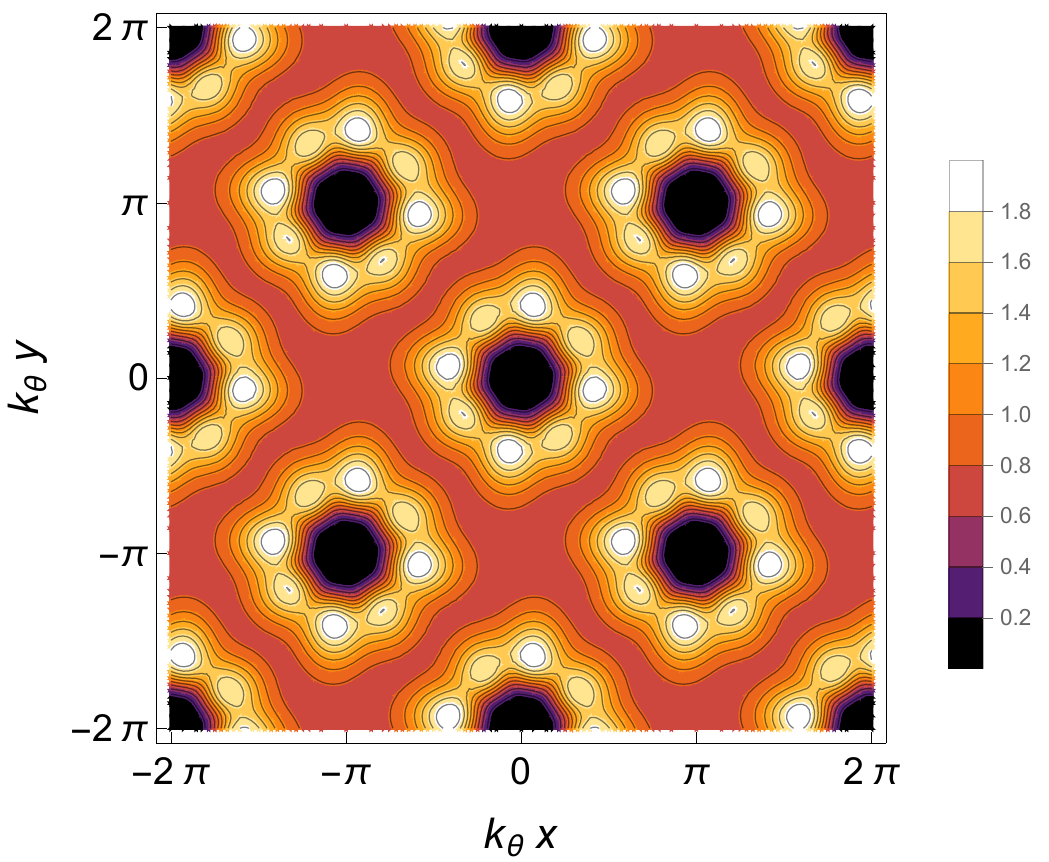}
\nonumber\\
(a) &(b)&(c)&(d)&(e)
\end{tabular}
\caption{
Density profiles of flat bands at $\Gamma^{\rm m}$ point in Fig. \ref{f:sp_m} (b) and listed in Table \ref{t:width},
where $k_\theta$ for the coordinates $x,y$ is restored.
Leftmost panel (a) shows the  top of the valence bands listed as 1 in Table \ref{t:width}, and to the right, 
the panels show
the second, third, fourth bands, and the rightmost panel (e) show the fifth band not listed in the table.
}
\label{f:den}
\end{center}
\end{figure*}

To elucidate the origin of these flat bands, 
we show in Fig. \ref{f:den} the density profile of the flat bands in Fig. \ref{f:sp_m} (b) (and listed in Table \ref{t:width}). 
It turns out that near the band center, the wave functions are well localized, each of which forms the moir\'e 
$AA$-sublattice illustrated in Fig. \ref{f:moire}. 
Decreasing the energy and going away from the band center, the wave functions becomes wider, and at the energy 
$\varepsilon/(v_0k_\theta)\sim - m$, the wave function becomes extended, as shown in Fig. \ref{f:den} (e).
This suggests that the localization of the wave function causes the flat bands in the present system.

\section{Flat bands as localized Landau levels}\label{s:landau}

So far we have discussed that flat bands are formed in the mass gap even 
in the absence of a magnetic field other than $\pi$ flux. 
Here, the localization is a key to understand the flat bands, as discussed in the previous section.
To clarify these flat bands, we derive an effective local Hamiltonian valid within the mass gap.

First, we argue that the moir\'e potential $\alpha$ plays a role of a magnetic field.
The moir\'e potential in Eq. (\ref{Ham_re}) [see also Eq. (\ref{FinHamA})] is written as 
$\alpha_0\cos(\theta/2)v\sigma^0\tau^1+\alpha_0\sin(\theta/2)\sigma^3\tau^2
+2\alpha(\sin x\sigma^2-\sin y\sigma^1)\tau^2$ in unit of $v_0k_\theta$.
Therefore, diagonalization of $\tau^2$ by the  unitary transformation 
$\tau^2\leftrightarrow\tau^3$, $\tau^1\rightarrow-\tau^1$ leads to 
\begin{widetext}
\begin{alignat}1
{\cal H}&=v_0k_\theta\left(\begin{array}{cccc}
 m+\alpha_{0\rm s}v(\bm x) &-2i\partial-i\alpha\bar u(\bm x) &-\alpha_{0\rm c}v(\bm x) & 0\\ 
 -2i\bar\partial+i\alpha u(\bm x)&-m-\alpha_{0\rm s}v(\bm x) &0&-\alpha_{0\rm c}v(\bm x)\\
 -\alpha_{0\rm c}v(\bm x)& 0&m-\alpha_{0\rm s}v(\bm x) & -2i\partial +i\alpha \bar u(\bm x)\\
 0&-\alpha_{0\rm c}v(\bm x)&-2i\bar\partial -i\alpha u(\bm x)&-m+\alpha_{0\rm s}v(\bm x) \\
\end{array}\right),
\label{EffHam3}
\end{alignat}
\end{widetext}
where $\alpha_{0\rm s}=\alpha_0\sin(\theta/2)$ and $\alpha_{0\rm c}=\alpha_0\cos(\theta/2)$.
For reference, we derive the above Hamiltonian via the chiral basis convenient for the TBG in Appendix \ref{a:basis}.
Remarkably, $\alpha $-potential is completely incorporated into the kinetic term  
and serve as a (periodic) magnetic field: 
Let $-iD=-i\partial-e\bar A$ and $-i\bar D=-i\bar\partial -eA$ be covariant derivatives,
where $A=(A_x+iA_y)/2$.
Then, $[-2iD,-2i\bar D]=-2eB(\bm x)$ gives a magnetic field perpendicular to the $x$-$y$ plane.
In the present Hamiltonian (\ref{EffHam3}), we have 
$[-2i\partial\mp i\alpha \bar u,-2i\bar\partial\pm i\alpha u]=\pm2\alpha v$,
implying that the $AB$ staking potential $u(\bm x)$ and $AA$ staking potential $v(\bm x)$ serve, respectively,  as a vector potential and
a magnetic field. 
In spite of such an effective magnetic field induced by the moir\'e potentials,
TR invariance is preserved due to opposite effective charges of the doubled Dirac fermions.
The Hamiltonian (\ref{EffHam3}) also tells that when $\alpha_0=0$, two Dirac fermions are decoupled, and 
it is easy to prove that
each Hamiltonian has energies  bounded by  $\varepsilon^2\ge  (v_0k_\theta m)^2$, as advertised 
in Sec. \ref{s:numerical}.
This is the case in Fig. \ref{f:sp_a} (b).

\subsection{Landau levels around zero energy}

We have argued that the moir\'e potential serves as a periodically oscillating  magnetic field.
In order to further reveal  the nature of the flat bands found in Sec. \ref{s:numerical}, 
we assume that 
eigenstates around zero energy within the mass gap are dominated by the potential $|u(\bm x)|\ll m$
to make the kinetic energy as small as possible.
This occurs at $x\sim l\pi$ and $y\sim n\pi$, where $l,n$ are integers. Around this point, $u(\bm x)\sim0$, 
whereas $v(\bm x)\sim 4$ for $l+n=$ even ($AA$ stack point) and $v(\bm x)\sim 0$ for $l+n=$ odd ($AB$ stack point).
As discussed above, the latter case cannot yield any states within the mass gap.
Thus we can use the moir\'e potential $u(\bm x)$ expanded up to the linear order of $\bm x$ valid around 
$z\sim z_{ln}\equiv (l+in)\pi$ with $l+n=$ even.
To be concrete, we have
\begin{alignat}1
&u(\bm x)\sim 2
(-)^{l} (z-z_{ln}) ,
\nonumber\\
&v(\bm x)\sim2
(-)^{l} (2-|z-z_{ln}|^2/2) .
\end{alignat}
Note here that we have used the coordinates rescaled by $k_\theta$, 
as introduced in Sec. \ref{s:symmetry}.
At each point labeled by $z_{ln}$, we can define an effective local Hamiltonian around zero energy. 
As the sign of $\alpha$ is always fixed to be positive by a gauge transformation, 
we assume $\alpha >0$ without loss of generality. Then, 
let us define the operators
\begin{alignat}1
a&=\frac{-i}{\sqrt{2\alpha }}\left[\partial+\alpha (\bar z-\bar z_{ln})\right],
\nonumber\\
a^\dagger&=\frac{-i}{\sqrt{2\alpha}}\left[\bar\partial-\alpha ( z-z_{ln})\right],
\nonumber\\
b&=\frac{-i}{\sqrt{2\alpha}}\left[\bar \partial+\alpha  (z-z_{ln})\right],
\nonumber\\
b^\dagger&=\frac{-i}{\sqrt{2\alpha}}\left[\partial-\alpha  (\bar z-\bar z_{ln})\right],
\label{ComLoc}
\end{alignat}
which satisfy
\begin{alignat}1
[a,a^\dagger]=[b,b^\dagger]=1, \quad [a,b]=[a,b^\dagger]=0.
\end{alignat}
Note that $a$ and $b$ operators have opposite charges which ensures the TR symmetry 
when $m=0$, as mentioned above.
In the case of even $l$, the local Hamiltonian valid near the point $z_{ln}$
can be expressed as
\begin{alignat}1
{\cal H}&=v_0k_\theta\left(\begin{array}{cccc}
 M_-&2\sqrt{2\alpha }a &-\alpha_{0\rm c}V& 0\\ 
 2\sqrt{2\alpha }a^\dagger&-M_- &0&-\alpha_{0\rm c}V\\
 -\alpha_{0\rm c}V& 0&M_+ & 2\sqrt{2\alpha }b^\dagger\\
 0&-\alpha_{0\rm c}V&2\sqrt{2\alpha }b&-M_+ \\
\end{array}\right),
\label{LocHam}
\end{alignat}
where $M_\pm=m\pm \alpha_{0\rm s}V$ and
\begin{alignat}1
V&=
4-\frac{1}{2\alpha}(a-b^\dagger)(a^\dagger-b)
\end{alignat}
In the case of odd $l$, the creation and annihilation operators are exchanged in the Hamiltonian (\ref{LocHam}), 
$a\rightarrow b^\dagger $, $b\rightarrow a^\dagger$ and $V\rightarrow -V$.

The Hamiltonian (\ref{LocHam}) would describe states localized at the moir\'e site $z_{ln}$,
if it allows eigenstates near zero energy within the mass gap. 
Conventionally, if we regard $a$ and $a^\dagger$ as operators describing the cyclotron motion,
then $b$ and $b^\dagger$ are the operators describing the guiding center.
However, in the present system, it might be suitable to interpret them as doubled fermions with opposite charges 
associated with two layers.
Since there are many $AA$ stacking points $z_{ln}$,
these state are degenerate over the $x$-$y$ plane and form flat bands.
However, this is far from trivial, since the Landau levels of the Hamiltonian (\ref{LocHam}) is 
$(\varepsilon/v_0k_\theta)^2=m^2+8\alpha  n>m^2$ when $\alpha_0=0$. 
In Table \ref{t:width}, numerically computed energies of the Hamiltonian (\ref{LocHam}) are listed. 
It turns out that the Hamiltonian (\ref{LocHam}) yields eigenstates around zero energy which 
indeed reproduce those of the full system.  
We therefore conclude that this is the origin of the flat bands in the present system.
These flat bands may be referred to as moir\'e Landau levels, since they occur due to a uniform (effective) magnetic field
originated from the moir\'e potential, although the mechanism of the degeneracies are quite different from 
the conventional Landau levels of the quantum Hall effect.

\section{Summary and discussion}\label{s:summary}
We studied  a  moir\'e system with $C_4$ symmetry.
As a lattice model, we considered twisted bilayer system composed of the conventional 
square lattice with $\pi$ flux per plaquette, which is known to yield linear dispersion around zero energy
like the graphene. 
According to BM, we derived a doubled massive Dirac Hamiltonian as an effective theory of moir\'e system
and studied its spectrum.
In the case of TBG with $C_3$ symmetry, the massless Dirac model 
with chiral symmetry allows perfect flat bands at magic angles, 
whereas the twisted $\pi$-flux system with $C_4$ symmetry,
the massless model never allows flat bands due to TR symmetry. 
Rather, the massive model shows Landau-level-like flat bands 
even in the absence of a magnetic field other than $\pi$ flux. 
This is due to the fact that the $ C_4$ symmetric moir\'e potential serves as a periodic magnetic field, 
and within the mass gap, such a magnetic field can be regarded as uniform.
Even in the presence of such an effective magnetic field, the doubled Dirac fermions have the opposite 
effective charges, which ensures the TR invariance of the total system in the massless case.
These flat bands may be called moir\'e Landau levels, since they are associated with a uniform magnetic field,
but their degeneracies are due to the periodicity of the moir\'e potential.

To elucidate the bilayer Dirac system with $C_4$ symmetry, 
we resorted to the $\pi$-flux model, since it is a minimal two-band system. 
While this lattice model is very specific, the effective Dirac model
in the continuum limit is rather universal in that the interlayer potential can be derived 
by $C_4$ symmetry properties only.  
Thus, as more realistic models,  a $C_4$ symmetric Dirac semimetal \cite{Young:2015uf} 
may be candidates for the moir\'e Landau levels proposed in this paper.
Since they appear in gapped systems,  symmetry-breaking perturbations giving a mass gap are needed.
This model includes the spin-orbit coupling and its Dirac points are protected 
by nonsymmorphic symmetries, so that effects of spins and 
nonsymmorphic symmetries, if not broken, for the effective Dirac continuum model of the twisted systems
may be an interesting future problem.



\begin{figure}[h]
\begin{center}
\begin{tabular}{cc}
\includegraphics[width=0.4\linewidth]{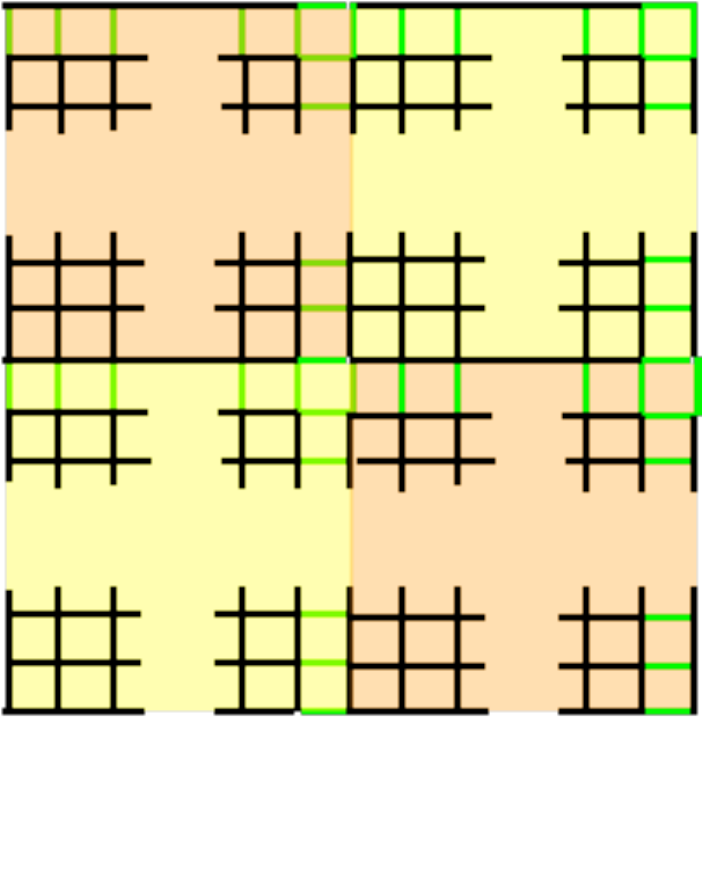}
&
\includegraphics[width=0.55\linewidth]{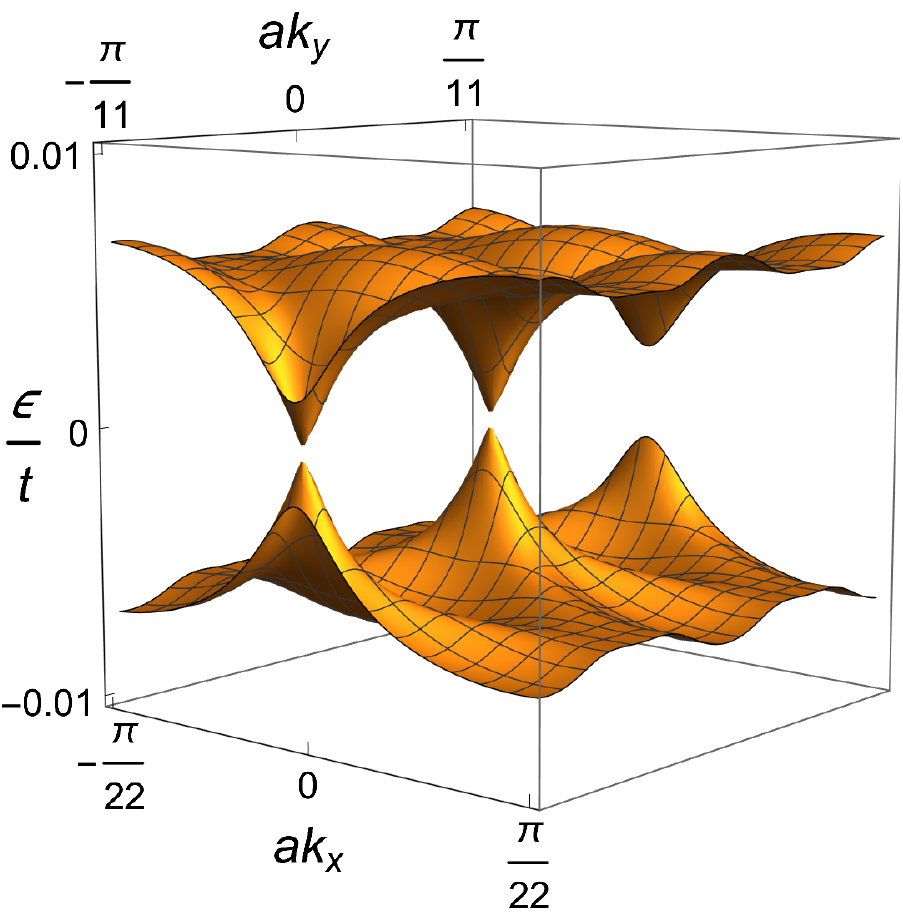}\end{tabular}
\caption{
Left panel shows the square lattice with uniform $\phi=\pi/n^2$ flux per mini-plaquette.
Each colored region (yellow or orange) is penetrated by uniform $\pi$ flux, provided that it includes $n^2$ plaquettes.
The black and green bonds stand for the hoppings $t$ and $t'$, respectively.
Right panel shows the spectra of two bands around zero energy in the case of
$t'=0.2t$ and $n=11$. 
}
\label{f:ef}
\end{center}
\end{figure}

Finally, let us discuss a model which has  more intimate relationship with the $\pi$-flux model.
In general, a uniform magnetic field directly giving $\pi$ flux per plaquette is too large  to be realized in experiments.
Toward the experimental realization of the moir\'e Landau levels using models associated with the $\pi$-flux model,
one may consider the possibility of using the moir\'e technologies.
Namely,  taking account of the fact that moir\'e pattern gives an enlarged but the same periodic structure 
as the original crystalline structure, 
one can tune the square moir\'e lattice to have $\pi$ flux per moir\'e plaquette using two layers with a weak magnetic field.
If one consider the twisted bilayer model of such moir\'e systems, the moir\'e of moir\'e  pattern 
(or simply twisted multilayer system) would yield the moir\'e Landau flat bands presented in this paper, 
although it is beyond the scope of this paper.
Alternatively, we shall show that if a long-periodic structure given  on the lattice
includes $\pi$ flux per each unit cell, a Dirac-like dispersion 
is induced likewise the $\pi$-flux model,  even though the magnetic flux per mini-plaquette is small.

In Fig. \ref{f:ef}, a simple square lattice is presented,
on which a particle is hopping between nearest-neighbor sites
with strength $t$ and $t'$ denoted by black and green lines, respectively, 
where $t'$ defines an enlarged unit plaquette marked by yellow and orange. 
If one chooses $\phi=\pi/n^2$, the unit cell includes just $\pi$ flux. 
This model exhibits many Landau levels due to a magnetic field, which are almost flat compared with 
the energy $\sim 2t$. However, in more fine energy scale $\sim 2t/n^2$, the Landau levels are dispersive.
Indeed, the spectrum in Fig. \ref{f:ef} shows doubled 
Dirac dispersions in the Brillouin zone, even though the flux in each mini-plaquette is $\pi/11^2$.
One can expect that the twisted bilayer system of such a model could show the moir\'e Landau levels presented 
in this paper. 
We also believe that many other moir\'e systems composed of models with 
enlarged lattice structures, including moir\'e of moir\'e system mentioned above, 
have possibility of the experimental realization of the moir\'e Landau flat bands presented in this paper. 

As another example with $C_4$ symmetry, a twisted bilayer system composed of $d\pm id'$-wave superconductors 
have been studies in Ref. \cite{Can:2021vu}. From the symmetry argument in Sec. \ref{s:IC_sym}, 
this model is also expected to show a similar properties to the present model at small twist angles. 
On the other hand,  since this model has Dirac nodes at different points in the Brillouin zone and 
the Brillouin zone is folded due to the Cooper pairing, 
interlayer potentials could be different from those of our model as derived in Sec.  \ref{s:IC_micro}.
This may be an interesting issue to be explored.

\acknowledgements
TF would like to thank T. Fujiwara for valuable discussions. 
This work was supported in part by Grants-in-Aid for Scientific Research Number  17H06138
from the Japan Society for the Promotion of Science.

\appendix

\section{Fourier transformation of the interlayer coupling}\label{a:fourier}

\begin{figure}[h]
\begin{center}
\begin{tabular}{c}
\includegraphics[width=0.9\linewidth]{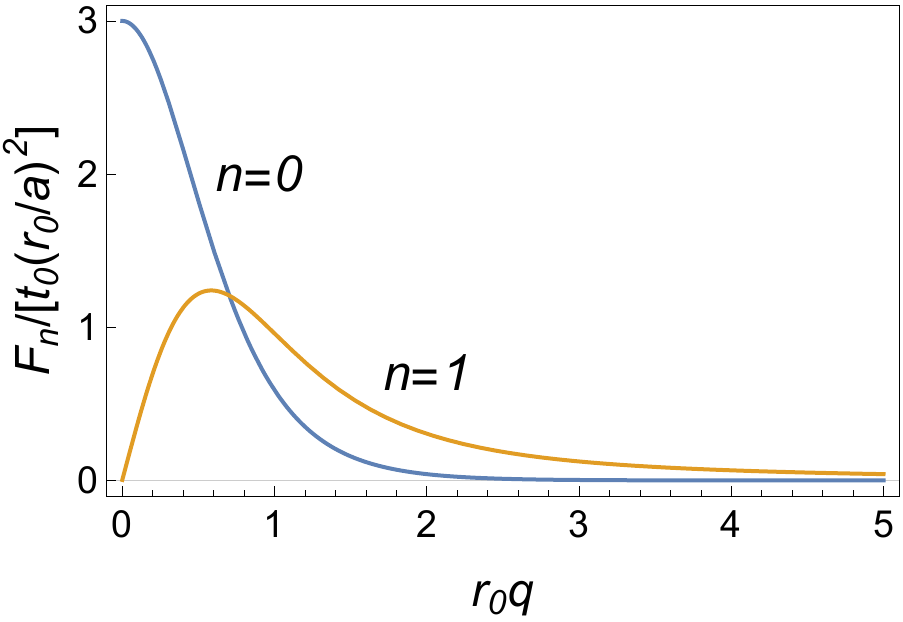}
\end{tabular}
\caption{
Fourier transformation $F_n(q)$ in Eq. (\ref{FouTra}) for $f(r)$ given by Eq. (\ref{IntCou}) with
$z_0/r_0=2$.
}
\label{f:fourier}
\end{center}
\end{figure}

When the effective  interlayer coupling for the continuum model is derived, the 
Fourier transformation is needed. To this end, let us consider the Fourier transformation of the function 
$f_n(\bm x)=f(r)e^{in(\theta+\theta_0)}$, 
\begin{alignat}1
F_n(\bm q)&=\int_{-\infty}^\infty \frac{d^2\bm x}{a^2}f_n(\bm x)e^{-i\bm q\cdot\bm x},
\end{alignat}
where $a$ is the lattice constant, introduced for later convenience.
Its inverse is
\begin{alignat}1
f_n(\bm x)=\int_{-\infty}^\infty\frac{d^2q}{(2\pi/a)^2}e^{i\bm q\cdot\bm x}F_n(\bm q).
\label{IntLayFou}
\end{alignat}
Introducing $\bm q=q(\cos\phi,\sin\phi)$, we have
\begin{alignat}1
F_n(\bm q)&=e^{in\theta_0}\frac{1}{a^2}\int_0^\infty drrf(r)\int_0^{2\pi}d\theta e^{i[n\theta-qr\cos(\theta-\phi)]}
\nonumber\\
&= e^{in\theta_0}e^{in(\phi-\pi/2)}\frac{2\pi}{a^2}\int_0^\infty drrf(r)J_n(qr)
\nonumber\\
&\equiv e^{in\phi}F_n(q),
\label{FouTra}
\end{alignat}
where $J_n(z)$ stands for the $n$th Bessel function.
Let us assume 
\begin{alignat}1
f(r)=t_0e^{-\left[\sqrt{r^2+z_0^2}-z_0\right]/r_0},
\label{IntCou}
\end{alignat}
where $t_0$ is the strength of the interlayer coupling when $r=0$ and 
$r_0$ is the correlation length of the interlayer coupling. We show in Fig. \ref{f:fourier} the real parts of the Fourier transformations
$F_n(q)$ ($n=0,1$). 
Since that $F_1(q)$ has a peak around $r_0q\sim0.6$,
$r_0$ should be $r_0>0.6a/\pi\sim0.2a$ for $F_n(q)$ to decrease rapidly for $q>k_{\rm M}=\pi/a$.

\section{Derivation of the interlayer coupling}\label{a:interlayer}

In this Appendix, we give a brief review of deriving the effective interlayer coupling for the continuum Dirac model
\cite{Bistritzer:2011ab,Moon:2013vv}.
The Fourier transformation of the fermion operators for the upper and lower layers are defined by
\begin{alignat}1
&c_{a,\bm k}=\sum_{\bm r}e^{i\bm k\cdot\bm r_a}c_{a,\bm r^+},
\nonumber\\
&\tilde c_{a,\tilde{\bm k}}=\sum_{\tilde{\bm r}}e^{i\bm k\cdot\tilde{\bm r}_a}\tilde c_{a,\bm r^-},
\end{alignat}
where $a=A,B$ and $\bm r_a\equiv\bm r+\bm\tau_a$ and likewise for $\tilde{\bm r}_a$.
It should be noted here that the above definitions are slightly different from those in Sec. \ref{s:lattice_single}.
Their inverse are
\begin{alignat}1
&c_{a,\bm r}=\int_{-\pi/a}^{\pi/a}\frac{d^2k}{(2\pi/a)^2}e^{i\bm k\cdot\bm r}c_{a,\bm k},
\nonumber\\
&\tilde c_{a,\tilde{\bm r}}=\int_{-\pi/a}^{\pi/a}\frac{d^2\tilde k}{(2\pi/a)^2}e^{i\tilde{\bm k}\cdot\tilde{\bm r}}\tilde c_{a,\tilde{\bm k}},
\end{alignat}

Using the Fourier transformation Eq. (\ref{IntLayFou}), 
the interlayer coupling (\ref{IntLay}) is then rewritten by 
\begin{widetext}
\begin{alignat}1
H_{ab}&
=\sum_{\bm r,\tilde{\bm r}}\int_{-\infty}^\infty\frac{d^2q}{(2\pi/a)^2}
\int_{-\pi/a}^{\pi/a}\frac{d^2k}{(2\pi/a)^2}\int_{-{\pi/a}}^{\pi/a}\frac{d^2\tilde{k}}{(2\pi/a)^2}
t_{ab}(\bm q)e^{i\bm q\cdot(\bm r_a-\tilde{\bm r}_b)}e^{-i\bm k\cdot\bm r_a}e^{i\tilde{\bm k}\cdot\tilde{\bm r}_b}
c_{a,\bm k}^\dagger \tilde c_{b,\tilde{\bm k}}.
\end{alignat}
Note that the sums over $\bm r$ and $\tilde{\bm r}$ yield periodic $\delta$-function,
$\sum_{\bm r}e^{i\bm k\cdot\bm r}= (2\pi/a)^2\sum_{\bm G}\delta(\bm k-\bm G)$.
Therefore, generic interlayer coupling can be denoted in momentum representation such that
\begin{alignat}1
H_{ab}
&=\sum_{\bm G,\tilde{\bm G}}\int_{-{\pi/a}}^{\pi/a}\frac{d^2k}{(2\pi/a)^2}
t_{ab}(\bm k+\bm G)e^{i\bm G\cdot\bm \tau_a}e^{-i\tilde{\bm G}\cdot\tilde{\bm\tau}_b}
c_{a,\bm k}^\dagger \tilde c_{b,\bm k+\bm G-\tilde{\bm G}}.
\end{alignat}
For the purpose of deriving the interlayer coupling around the Dirac point $\bm k\sim \bm k_{\rm M}$.
To this end, set $\bm k=\bm k_{\rm M}+\bm k'$.
We assume that the region of the integration over $\bm k'$ is so small that $t_{ab}(\bm k_{\rm M}+\bm k')$
is almost constant. Then
\begin{alignat}1
H_{ab}&\sim
\sum_{\bm G,\tilde{\bm G}}
t_{ab}(\bm k_{\rm M}+\bm G)e^{i\bm G\cdot\bm \tau_a}e^{-i\tilde{\bm G}\cdot\tilde{\bm\tau}_b}
\int\frac{d^2k'}{(2\pi/a)^2}
c_{a,\bm k_{\rm M}+\bm k'}^\dagger \tilde c_{b,\bm k_{\rm M}+ \bm k'+\bm G-\tilde{\bm G}}.
\label{AppH}
\end{alignat}
We now define the fermion operators $ c_{a,\bm x}$ and $\tilde c_{a,\bm x}$ in the continuum space,
\begin{alignat}1
&c_{a,\bm k_{\rm M}+\bm k'}=\int_{-\infty}^\infty\frac{d^2x}{a^2}e^{-i\bm k'\cdot\bm x}c_{a,\bm x}
\nonumber\\
&\tilde c_{a,\bm k_{\rm M}+\bm k'}=\int_{-\infty}^\infty\frac{d^2x}{a^2}e^{-i\bm k'\cdot\bm x}\tilde c_{a,\bm x},
\end{alignat}
and extend the integration of $\bm k'$ in Eq. (\ref{AppH}) to infinity in the spirit of the continuum limit. 
Then,  we reach
\begin{alignat}1
H_{ab}&=\int_{-\infty}^\infty\frac{d^2x}{a^2}\sum_{\bm G,\tilde{\bm G}}
t_{ab}(\bm k_{\rm M}+\bm G)e^{i\bm G\cdot\bm \tau_a}e^{-i\tilde{\bm G}\cdot\tilde{\bm\tau}_b}
e^{-i(\bm G-\tilde{\bm G})\cdot\bm x}
c_{a,\bm x}^\dagger \tilde c_{b,\bm x},
\end{alignat}
which corresponds to Eq. (\ref{IntCouGen1}) with Eq. (\ref{IntCouGen2}).

\section{Basis changes of the Hamiltonian}\label{a:basis}

In order to derive the Hamiltonian (\ref{EffHam3}), 
it may be convenient to switch to the chiral basis used in Ref. \cite{Tarnopolsky:2019aa}. 
The Hamiltonian (\ref{Ham_re}) with (\ref{FinHam}) is explicitly given by
\begin{alignat}1
{\cal H}&=v_0k_\theta\left(\begin{array}{cccc}
 m&-2i\partial& \alpha_0e^{-i\theta/2}v(\bm x)  & -\alpha \bar u(\bm x)\\ 
 -2i\bar\partial& -m&\alpha u(\bm x)&\alpha_0e^{i\theta/2} v(\bm x) \\
\alpha_0e^{i\theta/2} v(\bm x)&\alpha \bar u(\bm x)&m&-2i\partial\\
-\alpha u(\bm x) &\alpha_0e^{-i\theta/2}v(\bm x)&-2i\bar\partial &-m\\
\end{array}\right).
\label{FinHamA}
\end{alignat}
By applying the orthogonal transformation exchanging $2\leftrightarrow3$ species, we rewrite the Hamiltonian as follows;
\begin{alignat}1
{\cal H}&=v_0k_\theta\left(\begin{array}{cccc}
 m& \alpha_0e^{-i\theta/2}v(\bm x) &-2i\partial & -\alpha \bar u(\bm x)\\ 
 \alpha_0e^{i\theta/2} v(\bm x)& m&\alpha \bar u(\bm x)& -2i\partial\\
-2i\bar\partial&\alpha u(\bm x)&-m&\alpha_0e^{i\theta/2} v(\bm x)\\
-\alpha u(\bm x) &-2i\bar\partial &\alpha_0e^{-i\theta/2}v(\bm x)&-m\\
\end{array}\right).
\label{EffHam}
\end{alignat}
When $m=0$ and $\alpha_0=0$, chiral symmetry is manifest. 
It should be noted that $\alpha $-potential is included as 
$i\alpha  \left[u(\bm x)\sigma^2\tau^--\bar u(\bm x)\sigma^2\tau^+\right]$, 
implying that the unitary transformation $\sigma^2\leftrightarrow\sigma^3$, $\sigma^1\rightarrow-\sigma^1$
moves the $\alpha $-potential into the kinetic terms such that
\begin{alignat}1
{\cal H}&=v_0k_\theta\left(\begin{array}{cccc}
 m+\alpha_{0\rm s}v(\bm x)& -\alpha_{0\rm c}v(\bm x) &-2i\partial -i\alpha \bar u(\bm x)&0\\ 
 -\alpha_{0\rm c}v(\bm x)& m-\alpha_{0\rm s}v(\bm x)&0& -2i\partial+i\alpha \bar u(\bm x)\\
-2i\bar\partial+i\alpha u(\bm x)&0&-m-\alpha_{0\rm s}v(\bm x)&-\alpha_{0\rm c}v(\bm x)\\
 0&-2i\bar\partial -i\alpha u(\bm x)&-\alpha_{0\rm c}v(\bm x)&-m+\alpha_{0\rm s}v(\bm x)\\
\end{array}\right),
\end{alignat}
\end{widetext}
where $\alpha_{0\rm s}=\alpha_0\sin(\theta/2)$ and $\alpha_{0\rm c}=\alpha_0\cos(\theta/2)$.
Then,  we replace the $2\leftrightarrow 3$ species again, we obtain the Hamiltonian (\ref{EffHam3}).


\end{document}